\documentclass[twocolumn,aps,prc,tightenlines,floats,floatfix,eqsecnum,preprintnumbers]{revtex4}

\def\sp{\kern +3pt}
\def\sm{\kern -3pt}
\def\spQ{\kern +6pt}

\def\bea{\begin{eqnarray}}
\def\eea{\end{eqnarray}}

\def\sfrac#1#2{{\textstyle \frac{#1}{#2}}}

\def\intKP{\int_{\boldsymbol \kappa}}

\newcommand{\bra}[1]{\langle #1|}
\newcommand{\ket}[1]{|#1\rangle}

\def\be{\begin{equation}}
\def\ee{\end{equation}}
\def\ba{\begin{eqnarray}}
\def\ea{\end{eqnarray}}

\usepackage{graphics}
\usepackage{graphicx}
\usepackage{epsf}
\usepackage{amsmath}
\usepackage{amssymb}

\begin{document}

\phantom{0}
\vspace{-0.2in}
\hspace{5.5in}

\preprint{ADP-13-07/T827}

\vspace{-1in}

\title
{\bf Octet to decuplet electromagnetic transition
in a relativistic quark model}
\author{G.~Ramalho$^1$ and K.~Tsushima$^2$}
\vspace{-0.1in}

\affiliation{
$^1$CFTP, Instituto Superior T\'ecnico,
Universidade T\'ecnica de Lisboa,
Avenida Rovisco Pais, 1049-001 Lisboa, Portugal \vspace{-0.15in}}
\affiliation{
$^2$CSSM, School of Chemistry and Physics,
University of Adelaide, Adelaide SA 5005, Australia
}

\vspace{0.2in}
\date{\today}

\phantom{0}

\begin{abstract}
We study the octet to decuplet baryon electromagnetic transitions
using the covariant spectator quark model and predict
the transition magnetic dipole form factors for those involving
the strange baryons.
Utilizing $SU(3)$ symmetry,
the valence quark contributions
are supplemented by the pion cloud dressing based on
the one estimated in the
$\gamma^\ast N \to \Delta$ reaction.
Although the valence quark contributions
are dominant in general, the pion cloud effects
turn out to be very important
to describe the experimental data.
We also show that other mesons besides the pion
in particular the kaon, may be relevant for some
reactions such as $\gamma^\ast \Sigma^+ \to \Sigma^{*+}$,
based on our analysis for the radiative decay widths
of the strange decuplet baryons.
\end{abstract}

\vspace*{0.9in}  
\maketitle

\section{Introduction}

Low-lying baryons are classified into the
spin 1/2 octet and spin 3/2 decuplet
by quark models and quantum chromodynamincs (QCD).
The electromagnetic structure of the octet ($B$)
and decuplet ($B'$) baryons,
or the $\gamma^* B \to B'$ transition, can be characterized
by their electromagnetic form factors.
These form factors encode the microscopic
quark and gluon QCD substructure of the baryons,
but can also be represented in
terms of the effective degrees of freedom, such as
the baryon cores dressed by meson clouds.

Although there is abundant experimental information on the
nucleon electromagnetic structure
and the $\gamma^\ast N \to \Delta$ transition
form factors~\cite{Pascalutsa07,Burkert04a,NSTAR},
in particular, the studies of the other possible
octet to decuplet electromagnetic
transition form factors involving the strange baryons
(baryons with one or more strange quarks),
are nearly nonexistent.
(The data can be found
in Refs.~\cite{PDG,Keller11a,Keller11b,Taylor05,Molchanov04}.)
In the past, several theoretical studies
of the octet to decuplet electromagnetic
transitions were performed in nonrelativistic and relativistic
quark models~\cite{Lipkin73,Koniuk80,Darewych83,Kaxiras85,Warns,Sahoo95,Wagner98,Tiator04,Sharma10,Sharma13,Santopinto12},
Skyrme and soliton models~\cite{Schat96,Abada96,Haberichter97},
QCD sum rules~\cite{Wang09},
chiral perturbation theory~\cite{Butler93a,Butler93b,Arndt04},
large $N_c$ limit~\cite{Lebed11},
algebraic models of hadron structure~\cite{Bijker00},
and lattice QCD~\cite{Leinweber93}.
In  particular,
lattice QCD studies for the
$\gamma^* N \to \Delta$ reaction can be found
in Refs.~\cite{Alexandrou05,Alexandrou08}.

The study of the $\gamma^* B \to B'$ transition
is very important to understand
the role of the meson cloud.
For the $\gamma^* N \to \Delta$ reaction,
the meson cloud contributions are
shown to be crucial~\cite{Pascalutsa07,Burkert04a,Tiator04,NDelta,NDeltaD}.
Then, it is natural to investigate
the role of the meson cloud also for the other
octet to decuplet electromagnetic transitions.
In order to understand the role of the meson cloud quantitatively,
where the pion cloud is expected to be dominant,
the prerequisite is to understand the valence quark
contributions quantitatively.

For this purpose, we rely on the covariant spectator quark
model~\cite{NDelta,NDeltaD,OctetMedium,OctetFF,Omega,Octet,ExclusiveR},
since it was successful in the studies of the
electromagnetic structure of nucleon~\cite{Nucleon,Nucleon2,FixedAxis},
octet and decuplet baryons~\cite{OctetMedium,OctetFF,Octet,Omega,GE2Omega,DeltaDFF,DeltaDFF2,Deformation},
transition form factors of the reactions
$\gamma^\ast N \to \Delta(1232)$,
$\gamma^\ast N \to N^\ast (1440)$,
$\gamma^\ast N \to N^\ast (1535)$~\cite{NDelta,NDeltaD,Lattice,LatticeD,RoperS11},
and others~\cite{Delta1600,Various}.
We follow the formalism
developed in Refs.~\cite{OctetMedium,OctetFF} for the octet baryons,
and that in Ref.~\cite{Omega}
for the decuplet baryons.
In these works, the covariant spectator quark model
was extended from the $SU(2)$ to the $SU(3)$ scheme
for the lattice QCD regime, and
then extrapolated back to the physical regime.
We also follow closely the study made for
the $\gamma^* N \to \Delta$ transition
based on an S-state approach to describe the
nucleon and $\Delta$ systems~\cite{NDelta}
utilizing the $SU(3)$ meson-baryon coupling scheme.
However, as is well known, the contributions solely from the valence quarks
are insufficient to describe the observed
cross sections and the extracted form factors
for the $\gamma^* N \to \Delta$ reaction
(especially the magnetic dipole form factor $G_M^*$).
Therefore, explicit pion cloud effects
for the other $\gamma^* B \to B'$ reactions
involving the strange baryons should also be considered
by extending the $\gamma^* N \to \Delta$ treatment
based on an $SU(3)$ symmetry scheme, and this
is done in the present study.

Since we adopt an S-state approximation
for the octet and decuplet systems,
the contributions for the electric and Coulomb
quadrupole form factors will vanish,
and only the contributions for the magnetic dipole
form factor $G_M^\ast$ will survive.
Although this is an approximation, 
it is justified, 
since the electric and Coulomb form factors
are known to be small compared to $G_M^\ast$ 
in the $\gamma^\ast N \to \Delta$ 
reaction~\cite{Pascalutsa07,Burkert04a,NSTAR,NDeltaD}. 
Then, in this article we will focus 
on the magnetic dipole form factor.

This article is organized as follows:
In Sec.~\ref{secSQM} we present
the covariant spectator quark model,
including the parametrizations for the
octet and decuplet baryon wave functions
and the quark electromagnetic current.
Results of the form factors, both for the valence quark
and pion cloud contributions, are presented in Sec.~\ref{secFF}.
Section~\ref{secResults} is devoted to the final results and discussions.
Summary and conclusions are given in Sec.~\ref{secConclusions}.

\section{Covariant spectator quark model}
\label{secSQM}

The covariant spectator quark model was derived from the
covariant spectator theory~\cite{Gross}.
In the model a baryon is described as
a three-constituent quark system, where one quark is free
to interact with the electromagnetic fields and a pair
of noninteracting quarks is treated as a single on-mass-shell
spectator particle (diquark) with an effective
mass $m_D$~\cite{Omega,Nucleon,Nucleon2}.
The quark current is parametrized based
on a vector meson dominance mechanism as explained
in detail in Refs.~\cite{Omega,Nucleon,OctetFF,OctetMedium}.

The baryon wave function depends on
the baryon momentum $P$, the diquark momentum $k$,
the flavor indices and the spin projections, as will be shown later.
The wave function is constructed conveniently
by the symmetrized states of the diquark
pair (12), and the off-mass-shell quark 3.
The transition current can be calculated
in terms of the quark-3 states.
To obtain the final, total contribution,
we may multiply by a factor of 3
the current associated with the quark 3.

The covariant spectator quark model was also
generalized to the lattice QCD regime
with heavy pions, where the meson cloud effects
are expected to be very small~\cite{Lattice,LatticeD,Omega}.
The fact that the same parametrization of the model
holds for both the physical and the lattice QCD regimes
gives us some confidence that the
valence quark contributions calculated
in the model are well under control.

Next, we will present the wave functions
of the octet and decuplet baryons in the
covariant spectator quark model.
We will use $M_B$ and $M_{B'}$ for the octet and
decuplet baryon masses, respectively, and
$M_N$ for the nucleon mass.

\subsection{Octet baryon wave functions}

In general the octet baryon wave function (spin 1/2)
in an S-state for the quark-diquark system 
can be written as~\cite{OctetFF,Octet}
\be
\Psi_B(P,k) =
\frac{1}{\sqrt{2}} \left\{
\phi_S^0 \ket{M_A} +
\phi_S^1 \ket{M_S}
\right\} \psi_B(P,k),
\label{eqPsiB}
\ee
where $\ket{M_A}$ and $\ket{M_S}$ are respectively the flavor
antisymmetric and symmetric wave functions,
$\phi_S^X$ ($X=0,1$) are the spin (0 and 1) wave functions,
and $\psi_B$ is the octet baryon $B$ radial wave function
to be defined shortly.
Spin projection indices
are suppressed for simplicity.
The explicit expressions for the all octet baryon members
are presented in Table~\ref{tableOctet}.
The spin wave functions are given by
\ba
& &
\phi_S^0 \left(+\sfrac{1}{2}\right)=
\frac{1}{\sqrt{2}}
\left( \uparrow \downarrow - \downarrow  \uparrow  \right) \uparrow, \\
& &
\phi_S^1
\left(+\sfrac{1}{2}\right)
= -
\frac{1}{\sqrt{6}}
\left[
\left( \uparrow \downarrow + \downarrow  \uparrow \right)
\uparrow - 2 \uparrow \uparrow \downarrow \right],
\ea
and
\ba
& &
\phi_S^0
\left(-\sfrac{1}{2}\right)=
\frac{1}{\sqrt{2}}
\left( \uparrow \downarrow - \downarrow  \uparrow  \right) \downarrow, \\
& &
\phi_S^1
\left(-\sfrac{1}{2}\right)=
\frac{1}{\sqrt{6}}
\left[
\left( \uparrow \downarrow + \downarrow  \uparrow \right)
\downarrow - 2 \downarrow \downarrow \uparrow \right].
\ea
This nonrelativistic structure is generalized
to a relativistic form in the covariant spectator quark model
\cite{Nucleon,NDelta}
\ba
\phi_S^0&=& u(P), \nonumber \\
\phi_S^1&=& -\varepsilon_\lambda^{\alpha \ast} (P)
U^\alpha(P),
\label{eqPhiS}
\ea
where
\be
U^\alpha(P)=
\frac{1}{\sqrt{3}}
\gamma_5 \left(
\gamma^\alpha-\frac{P^\alpha}{M_B}
\right) u(P).
\ee
In the above, $u(P)$
represents the Dirac spinor of the octet baryon $B$
with momentum $P$ and spin projection $s$,
and $\varepsilon_\lambda(P)$ ($\lambda=0,\pm 1$) the
diquark polarization vector in the fixed-axis
representation~\cite{Nucleon,FixedAxis}.
The spin projection is suppressed
in the Dirac spinors and $U^\alpha$ for simplicity.

The radial wave function $\psi_{B}$ is defined
in terms of the dimensionless variable $\chi_{_B}$
\ba
\chi_{_B}= \frac{(M_B-m_D)^2-(P-k)^2}{M_B m_D},
\label{eqCHI}
\ea
where $m_D$ is the diquark mass.
Using the formalism of Refs.~\cite{OctetMedium,OctetFF}
we write $\psi_{B}$ for $B=N,\Lambda,\Sigma,\Xi$
\ba
& &
\psi_{_N}(P,k)= \frac{N_N}{m_D(\beta_1+ \chi_{_N})(\beta_2+ \chi_{_N})},
\label{eqPsiN}
\\
& &
\psi_{_\Lambda}(P,k)= \frac{N_\Lambda}{m_D(\beta_1+ \chi_{_\Lambda})(\beta_3+ \chi_{_\Lambda})}, \\
& &
\psi_{_\Sigma}(P,k)= \frac{N_\Sigma}{m_D(\beta_1+ \chi_{_\Sigma})(\beta_3+ \chi_{_\Sigma})}, \\
& &
\psi_{_\Xi}(P,k)= \frac{N_\Xi}{m_D(\beta_1+ \chi_{_\Xi})(\beta_4+ \chi_{_\Xi})},
\label{eqPsiX}
\ea
where $N_B$ are the normalization constants
and $\beta_i$ ($i=1,2,3,4$)  are the momentum
range parameters in units of $m_D$.
We use the parameters determined in Ref.~\cite{OctetMedium},
namely, $\beta_1=0.0532$, $\beta_2=0.809$, $\beta_3=0.603$,
and $\beta_4=0.381$, in which we obtain natural
order for the size of the baryon cores~\cite{OctetMedium}.

The octet baryon masses can be described considering
the pion-baryon $SU(3)$ couplings.
We use the experimental baryon mass values:
$M_N= 0.939$ GeV, $M_\Lambda=1.116$ GeV, $M_\Sigma=1.192$ GeV
and $M_\Xi=1.318$ GeV.
The mass of the baryon $B$ in the octet can be
represented as $M_B= M_{0B} + \Sigma_0(M_B)$,
where $M_{0B}$ is a constant, and $\Sigma_0(M_B)$
the self-energy at the pole position,
which differs for the octet isomultiplet
($N,\Lambda,\Sigma$ and $\Xi$)~\cite{Octet}.
The self-energy is evaluated
neglecting the diagrams with heavy mesons
in the first approximation, and can be
expressed as $\Sigma_0= G_{0B} {\cal B}_0$,
where $G_{0B}$ is a factor depending on the
coupling of pion with the baryon $B$, and
${\cal B}_0$ is the value of the Feynman integral
(with the coupling constants removed) with the mass $M_B$.
The octet baryon masses can be reproduced
with an accuracy better than 7\% for the $SU(6)$ value
$\alpha \equiv D/(F+D) = 0.6$ with
$M_0= 1.342$ GeV and ${\cal B}_0= -0.127$ GeV~\cite{OctetMedium}.
The details of the $SU(3)$ couplings and coefficients $G_{0B}$,
are presented in Ref.~\cite{Octet}.

\begin{table*}[t]
\begin{center}
\begin{tabular}{l c c c}
\hline
\hline
$B$   & $\ket{M_S}$  & &  $\ket{M_A}$  \\
\hline
$p$     &   $\sfrac{1}{\sqrt{6}} \left[
        (ud + du) u - 2 uu d \right]$ & &
        $\sfrac{1}{\sqrt{2}} (ud -du) u$  \\
$n$     &  $-\sfrac{1}{\sqrt{6}} \left[
         (ud + du) d - 2 ddu \right]$ & &
         $\sfrac{1}{\sqrt{2}} (ud -du) d$ \\
\hline
$\Lambda^0$ &
$\sfrac{1}{2}
\left[ (ds+sd)u - (us+su)d
\right]$
& &
$\sfrac{1}{\sqrt{12}}
\left[
(sd-ds)u - (su-us)d +2(du-du)s
\right]$
\\
\hline
$\Sigma^+$  & $\sfrac{1}{\sqrt{6}} \left[(us + su) u - 2 uu s \right]$ &
            &  $\sfrac{1}{\sqrt{2}} (us -su) u $ \\
$\Sigma^0$ &
$\sfrac{1}{\sqrt{12}}
\left[
(sd+ds)u +(su+us)d -2(ud+du)s
\right]$
& &
$\sfrac{1}{2}
\left[ (ds-sd)u - (us-su)d
\right]$ \\
$\Sigma^-$ & $\sfrac{1}{\sqrt{6}}\left[ (sd + ds) d - 2 dd s \right]$ & &
              $\sfrac{1}{\sqrt{2}} (ds -sd) d$ \\
\hline
$\Xi^0$ & $-\sfrac{1}{\sqrt{6}} \left[(us + su) s - 2 ss u\right]$ & &
          $\sfrac{1}{\sqrt{2}} (us -su) s$ \\
$\Xi^-$ & $-\sfrac{1}{\sqrt{6}} \left[(ds + sd) s - 2 ss d\right]$ & &
          $\sfrac{1}{\sqrt{2}} (ds -sd) s$  \\
\hline
\hline
\end{tabular}
\end{center}
\caption{Flavor wave functions of the octet baryons~\cite{OctetFF,Octet}.}
\label{tableOctet}
\end{table*}

\subsection{Decuplet baryon wave functions}

We write down the decuplet baryon wave functions
in the S-state approximation for
the quark-diquark system~\cite{Omega}:
\ba
\Psi_{B'}(P,k)=
- \psi_{B'}(P,k) \varepsilon_\lambda^{\alpha *}(P) u_\alpha(P) \ket{B'},
\label{eqPsiBP}
\ea
where $\ket{B'}$ is the flavor state,
$u_\alpha$ is the Rarita-Schwinger vector-spinor and
$\varepsilon_\lambda (P)$ is
the diquark polarization vector in the decuplet baryon $B'$~\cite{FixedAxis}.
The explicit expressions are presented in Table~\ref{tableDec}.
The decuplet baryon radial
wave functions $\psi_{B'}$, for
$B'= \Delta,\Sigma^*,\Xi^*,\Omega$,
are given by~\cite{Omega}:
\ba
& &
\psi_{\Delta}(P,k) =
\frac{N_{\Delta}}{m_D
(\alpha_1 + \chi_{_{\Delta}})^3}, \\
& &
\psi_{\Sigma^*}(P,k) =
\frac{N_{\Sigma^*}}{m_D(\alpha_1 + \chi_{_{\Sigma^*}})^2
(\alpha_2 + \chi_{_{\Sigma^*}})}, \\
& &
\psi_{\Xi^*}(P,k) =
\frac{N_{\Xi^*}}{m_D(\alpha_1 + \chi_{_{\Xi^*}})
(\alpha_2 + \chi_{_{\Xi^*}})^2}, \\
& &
\psi_{\Omega}(P,k) =
\frac{N_{\Omega}}{m_D (\alpha_2 + \chi_{_{\Omega}})^3},
\ea
where $N_{B'}$ are the normalization constants and
$\chi_{B'}$ is given by Eq.~(\ref{eqCHI}) with
$M_B$ replaced by $M_{B'}$.
The wave function $\Psi_\Omega$ is also given
for completeness.
We use the parameters in Ref.~\cite{Omega},
$\alpha_1=0.3366$ and $\alpha_2=0.1630$.
A remark about the determination of these
parameters is in order.
The parameter $\alpha_1$ was determined in Refs.~\cite{NDeltaD,LatticeD}
using a model with the dominant S-state contribution
and very small D-state corrections,
utilizing physical and lattice QCD data
for the $\gamma^* N \to \Delta$ reaction.
The same value of $\alpha_1$ was
used in Ref.~\cite{Omega}.
In that work the lattice data
for the decuplet electromagnetic form factors~\cite{Boinepalli09}
were used to calibrate the value of $\alpha_2$,
neglecting the effects of the D-states.
This is justified by the smallness
of the D-state contributions, observed
previously in the $\gamma^* N \to \Delta$ transition
(smaller than 1\% in the $\Delta$ wave function)~\cite{LatticeD}.

As for the decuplet baryon masses, they can also
be described taking into
account the self-energy corrections
and the $SU(3)$ pion-baryon couplings.
In this case we write $M_{B'}=M_{0B'} + \Sigma_0^*(M_{B'})$
with $\Sigma_0^*= G_{1B'} {\cal B}_1+  G_{2B'} {\cal B}_2$,
where the terms in $G_{1B'}$ and $G_{2B '}$
are, respectively, associated with the intermediate states
of the octet and decuplet baryons, and depend on the baryon flavors.
${\cal B}_1$ and ${\cal B}_2$ are the Feynman integrals,
respectively for an intermediate baryon of the
octet and decuplet multiplet.
See Appendix~\ref{appMasses} for details.
We can reproduce the decuplet masses,
$M_\Delta=1.232$ GeV, $M_{\Sigma^*}= 1.385$ GeV,
$M_{\Xi^*}= 1.533$ GeV
and $M_\Omega= 1.672$ GeV, with a precision better than 0.1\%,
using
$M_{0B'}=1.672$ GeV, ${\cal B}_1=-0.544$ GeV and
${\cal B}_2=-0.266$ GeV.

\subsection{Transition current}

The electromagnetic
transition current $J^\mu$,
associated with the transition $\gamma^\ast B \to B'$,
can be written
in the relativistic impulse
approximation~\cite{Nucleon,Nucleon2,NDelta},
\ba
J^\mu \;= \; 3 \sum_{\lambda}
\int_k \overline \Psi_{B'}(P_+,k)j_q^\mu(q) \Psi_B (P_-,k),
\label{eqJB0}
\ea
where $P_+$ ($P_-$) is the final (initial)
baryon momentum,
$k$ the momentum of the on-shell diquark
and
$j_q^\mu(q)$ is the quark current,
depending on the transferred momentum $q= P_+ - P_-$
and on the quark flavor index ($u,d$ or $s$).
We represent the electromagnetic current
in units of the proton charge $e$.
Note the sum in the diquark polarization states ($\lambda=0,\pm 1$).
As mentioned already, the factor 3 takes into account
the sum in the quarks based on the wave function symmetry.
The integral symbol represents,
\be
\int_k= \int \frac{d^3 {\bf k}}{2 E_D(2\pi)^3},
\ee
with $E_D=\sqrt{m_D^2+ {\bf k}^2}$.

\subsection{Quark current}

The quark current $j_q^\mu$ effectively parametrizes the
constituent quark electromagnetic structure, and thus
includes the effects due to the gluon and meson dressing.

The operator $j_q^\mu$ has the
generic structure \cite{Nucleon,Omega,OctetMedium,Sanctis07},
\be
j_q^\mu(q) = j_1
\left(\gamma^\mu - \frac{{\not \! q}q^\mu}{q^2}
\right)
+ j_2 \frac{i \sigma^{\mu \nu} q_\nu}{2 M_N},
\label{eqJq}
\ee
where $M_N$ is the nucleon mass,
$j_i$ ($i=1,2$) are flavor operators acting on the
third quark in the $\ket{M_A}$ or $\ket{M_S}$ wave functions.
For the quark current we use,
\be
j_i=
\sfrac{1}{6} f_{i+} \lambda_0
+  \sfrac{1}{2}f_{i-} \lambda_3
+ \sfrac{1}{6} f_{i0} \lambda_s, \hspace{.2cm} (i=1,2),
\label{eqJi}
\ee
where
\ba
&\lambda_0=\left(\begin{array}{ccc} 1&0 &0\cr 0 & 1 & 0 \cr
0 & 0 & 0 \cr
\end{array}\right), \hspace{.3cm}
&\lambda_3=\left(\begin{array}{ccc} 1& 0 &0\cr 0 & -1 & 0 \cr
0 & 0 & 0 \cr
\end{array}\right),
\label{eqL1L3} \\
&\lambda_s \equiv \left(\begin{array}{ccc} 0&0 &0\cr 0 & 0 & 0 \cr
0 & 0 & -2 \cr
\end{array}
\right),
\ea
are the flavor matrices.
These operators act on
the quark wave function in flavor space
$q^T= (u\, d\, s)$.
The functions $f_{i\,\pm}(Q^2)$ with $Q^2=-q^2$
are the quark form factors
(see Ref.~\cite{OctetMedium} for details) and are normalized
as $f_{1\, \pm}(0)=1$,
$f_{2+}(0)=\kappa_+$,
$f_{2+}(0)=\kappa_-$
and $f_{20}(0)=\kappa_0$.
We can represent the isoscalar ($\kappa_+$),
isovector ($\kappa_-$) and $\kappa_0$ in terms of 
the quark $q=u,d,s$ anomalous magnetic moments,
defining $\kappa_q$ by $e_q \kappa_q \equiv j_2(0)$,
where $e_q$ is the quark charge. 
One obtains then
$\kappa_+ = 2\kappa_u -\kappa_d$,
$\kappa_- = \sfrac{1}{3}(2 \kappa_u + \kappa_d)$,
and $\kappa_0=\kappa_s$.
The values of the quark anomalous magnetic moments
were fixed in the previous works as
$\kappa_u=1.711$, $\kappa_d=1.987$,
and $\kappa_s= 1.462$~\cite{OctetMedium,Omega}.

Note that, the  values for the
quark anomalous magnetic moments $\kappa_q$,
defined according to Eq.~(\ref{eqJq}),
are expressed in units of nuclear magneton. 
In a naive conversion one can
use the constituent quark mass $m_q \approx M_N/3$,
which gives a factor $1/3$.
In the covariant spectator quark model
the anomalous magnetic moment takes into account
the internal structure of the constituent quark.
A simple estimate of the lowest-order effect
of the gluon to the electromagnetic vertex
gives $\kappa_q \simeq 1.5$~\cite{Nucleon}.
Therefore, deviations 
from the value 1.5 can
be interpreted as a consequence of the internal 
electromagnetic structure.
To compare our results with those of 
the quark anomalous magnetic moment usually
found in the literature, $\kappa_q^\prime$,
where the quark charge $e_q$ is
included in the definition,
we use $\kappa_q^\prime = \sfrac{1}{3} e_q \kappa_q$.
We obtain then
$\kappa_u^\prime= 0.380$, $\kappa_d^\prime= -0.221$
and $\kappa_s^\prime = -0.162$
[assuming $m_s=m_u,m_d$, according to $SU(3)$].

Our values for $\kappa_u^\prime$ and $\kappa_d^\prime$ are close to  
the results of others such as  
the  naive quark model~\cite{PDG2}, 
and calculations based on 
Dyson-Schwinger formalism~\cite{Chang11,Wilson12},
but are larger in absolute values  
than the other models such as, for instance, 
light-front constituent quark models~\cite{Cardarelli95} and  
the direct estimates of meson-cloud corrections~\cite{Ito95} 
($\kappa_u^\prime \simeq 0.1$; $\kappa_d^\prime = -0.15, -0.1$).

The inclusion of the term  $-{\not\! q} q^\mu / q^2$
in the quark current (\ref{eqJq}) is equivalent
to using the Landau prescription~\cite{Kelly98,Batiz98}
to the final electromagnetic current (\ref{eqJB0}).
The term restores current conservation but
does not affect the results of the observables~\cite{Kelly98}.
In the present study the correction
term gives no contribution to the transition current~(\ref{eqJB0}),
since the octet and decuplet states are orthogonal.

\begin{table}[t]
\begin{tabular}{l  c  c }
\hline
\hline
$B'$   & &  $\ket{B'}$      \\
\hline
\hline
$\Delta^{++}$  && $uuu$   \\
$\Delta^+$  && $\sfrac{1}{\sqrt{3}}\left[uud + udu + duu  \right]$
\\
$\Delta^0$  && $\sfrac{1}{\sqrt{3}}\left[ddu + dud + udd  \right]$
\\
$\Delta^-$  && $ddd$     \\
\hline 
$\Sigma^{\ast +}$ &&  $\sfrac{1}{\sqrt{3}} \left[uus  + usu + suu  \right]$
\\
$\Sigma^{\ast 0}$ &&
$\sfrac{1}{\sqrt{6}} \left[uds + dus +usd +  sud + dsu + sdu  \right]$
\\
$\Sigma^{\ast -}$ &&
$\sfrac{1}{\sqrt{3}} \left[dds + dsd + sdd  \right]$ \\
\hline 
$\Xi^{\ast 0}$ & $\;$& $\sfrac{1}{\sqrt{3}} \left[uss  + sus + ssu  \right]$
\\ 
$\Xi^{\ast -}$ && $\sfrac{1}{\sqrt{3}} \left[dss  + sds + ssd  \right]$
\\
\hline
$\Omega^-$     &&  $sss $
\\ 
\hline
\hline
\end{tabular}
\caption{Quark flavor wave functions $\left|B'\right>$ for the decuplet
baryons~\cite{Omega}.}
\label{tableDec}
\end{table}

\section{Electromagnetic form factors}
\label{secFF}

The $\gamma^* B \to B'$ transition current,
for the case of the initial $B$ spinor $u$
with momentum $P_-$ and the final $B'$ vector-spinor
$u_\beta$ with momentum $P_+$, can be
expressed as~\cite{NDelta},
\ba
& &
\hspace{-.5cm}
J^\mu = \nonumber\\
& &
\hspace{-.5cm}
\bar u_\beta (P_+)
\left[
G_1 q^\beta \gamma^\mu + G_2 q^\beta P^\mu +
G_3 q^\beta q^\mu - G_4 g^{\beta \mu}
\right] \! \gamma_5  u(P_-), \nonumber\\
\ea
where $P=\sfrac{1}{2}(P_+ + P_-)$.
For simplicity the $B$ and $B'$ spin
projections are suppressed.
In the above, $G_i$ ($i=1,2,3,4$) are
the octet to decuplet baryon transition form factors.
Only 3 of them are independent.
The current conservation leads to
the condition~\cite{NDelta}:
\ba
G_4= (M_{B'}+ M_B)G_1 + \frac{1}{2}(M_{B'}^2-M_B^2) G_2
- Q^2 G_3.
\nonumber \\
\ea

One can convert  the form factors $G_i$ ($i=1,2,3$) into
the multipole form factors defined by
Jones and Scadron~\cite{Jones73}:
\ba
G_M^* &=& K \left\{ \frac{}{} \!
\left[(3M_{B'}+M_B) (M_{B'}+M_B) + Q^2 \right] \frac{G_1}{M_{B'}}
\right.
\nonumber \\
& &
\left. + (M_{B'}^2-M_B^2) G_2 - 2 Q^2 G_3 \frac{}{} \right\},  \\
G_E^* &=&
K \left\{
(M_{B'}^2 - M_B^2 - Q^2 ) \frac{G_1}{M_{B'}}
\right.
\nonumber \\
& &
\left. + (M_{B'}^2-M_B^2) G_2 - 2 Q^2 G_3 \frac{}{} \right\},  \\
G_C^* &=&
K \left\{  \frac{}{}
4 M_{B'} G_1
+ (3 M_{B'}^2+ M_B + Q^2) G_2
\right.
\nonumber \\
& &
\left. + 2 (M_{B'}^2-M_B^2- Q^2) G_3 \frac{}{} \right\},
\ea
with
\be
K= \frac{M_B}{3(M_{B'} +M_B)}.
\ee
Hereafter, we use $G_X^*$  with $X=M,E,C$
to represent, respectively, the
magnetic dipole, electric quadrupole, and
Coulomb quadrupole form factors.

Next, we consider a decomposition,
$G_X^* = G_X^b + G_X^\pi$, where
$G_X^b$ is the contribution from the quark core
(valence quark contribution) and $G_X^\pi$
the pion cloud contribution.

\begin{table}[t]
\begin{tabular}{l  c  c }
\hline
\hline
    & &  $j_i^S$      \\
\hline
\hline
$\gamma^* p \to \Delta^+$  && $\sqrt{2}f_{i-}$     \\
$\gamma^* n \to \Delta^0$  &&
$\sqrt{2}f_{i-}$
\\[0.02in]
\hline 
$\gamma^* \Lambda \to \Sigma^{\ast 0}$ &&
$\sqrt{\frac{3}{2}} f_{i-}$
\\ 
\hline
$\gamma^* \Sigma^+ \to \Sigma^{\ast +}$  &&
$\frac{\sqrt{2}}{6}(f_{i+} + 3 f_{i-} + 2 f_{i0})$
\\
$\gamma^* \Sigma^0 \to \Sigma^{\ast 0}$  &&
$ \frac{\sqrt{2}}{6}(f_{i+} + 2 f_{i0})$
\\
$\gamma^* \Sigma^- \to \Sigma^{\ast -}$ &&
$\frac{\sqrt{2}}{6}(f_{i+} - 3 f_{i-} + 2 f_{i0})$  \\[0.02in]
\hline
$\gamma^* \Xi^0 \to \Xi^{\ast 0}$ &&
$\frac{\sqrt{2}}{6}(f_{i+} + 3 f_{i-} + 2 f_{i0})$
\\
$\gamma^* \Xi^- \to \Xi^{\ast -}$ &&
$\frac{\sqrt{2} }{6}(f_{i+} - 3 f_{i-} + 2 f_{i0})$
\\[0.02in]
\hline
\hline
\end{tabular}
\caption{
Coefficients $j_i^S$ ($i=1,2$) necessary
to calculate the valence quark contributions for the form factors.
}
\label{tableJS}
\end{table}

\subsection{Valence quark contributions}
\label{secValence}

Inserting the octet baryon $B$
and decuplet baryon $B'$ wave functions,
respectively, given by Eqs.~(\ref{eqPsiB}) and~(\ref{eqPsiBP})
into the transition current~(\ref{eqJB0}),
we calculate the valence quark contributions
for the current and form factors.

To perform the sum in the flavors associated
with the octet and decuplet baryons, we follow the procedure
given in Refs.~\cite{Octet,Omega}
with $\ket{B}$  and $\ket{B'}$
shown in Tables~\ref{tableOctet} and~\ref{tableDec},
and define:
\ba
{j_i^A}&\equiv&
3 \bra{B'} j_i \ket{M_A},\hspace{.4cm} (i=1,2), \\
{j_i^S}&\equiv&
3 \bra{B'} j_i \ket{M_S},\hspace{.4cm} (i=1,2).
\ea
The explicit results for $j_i^S$
are presented in Table~\ref{tableJS}.
As for $j_i^A$,
one has  $j_i^A \equiv 0$,
which reflects the orthogonality
between the spin-0 component of the octet baryon
and the spin-1 component of the decuplet baryon wave functions,
since the spin 3/2 states can have only
spin-1 diquarks.

Once the coefficients $j_i^S$ are determined,
we can calculate a factor $f_v$ 
and use the result of the current for the S-state 
approximation given by Ref.~\cite{NDelta}. 
One can write,
\ba
& &
\hspace{-.5cm}
J^\mu=
\frac{1}{\sqrt{3}} f_v {\cal I} \times \nonumber \\
& &
\hspace{-.5cm}
\bar u_\beta(P_+)
 \left[ 2 A M_{B'}\, q^\beta \gamma^\mu - 2 A q^\beta P^\mu
- A q^\beta q^\mu - g^{\beta \mu}
\right] \gamma_5 u(P_-), \nonumber \\
\ea
with  $A= \frac{2}{(M_{B'}+M_B)^2+ Q^2}$, and
\ba
& &{\cal I} (Q^2)= \int_k \psi_{B'}(P_+,k) \psi_B(P_-,k),
\label{eqInt}
\\
& &f_v(Q^2)=
\frac{1}{\sqrt{2}}\left\{
j_1^S(Q^2) +
\frac{M_{B'}+M_B}{2M_N}  j_2^S(Q^2) \right\},
\label{fv}
\nonumber \\
& &
\ea
where Eqs.~(\ref{eqInt}) and~(\ref{fv})
are respectively the overlap of the radial wave functions
and the symmetric flavor coefficient
(corresponding to the isovector coefficient in the
$\gamma^* N \to \Delta$ reaction).

From the relations above, one can derive,
\ba
& &G_1=  A M_{B'} G_4, \\
& &G_2= - A G_4, \\
& &G_3= - \frac{1}{2} A G_4,
\ea
with
\be
G_4= \frac{2}{\sqrt{3}} f_v {\cal I}.
\ee

The valence quark contributions are then given by:
\ba
G_M^b &=&
\frac{8}{3\sqrt{3}}
\frac{M_B}{M_{B'}+ M_B} f_v {\cal I},
\label{eqGMb}
\\
G_E^b &\equiv& 0,\\
G_C^b &\equiv& 0.
\label{eqGCb}
\ea
The result for $G_M^b$ depends on the details
of the baryon structure, namely the
radial wave functions $\psi_B$ and $\psi_{B'}$, through
the integral ${\cal I}$.
For $Q^2=0$ one can prove that
${\cal I} (0)\le 1$,
establishing the upper limit of $G_M^b(0)$ as
$\overline{G_M^b}(0)=\sfrac{8}{3\sqrt{3}} \frac{M_B}{M_{B'}+ M_B} f_v(0)$.
The results are given in Table~\ref{tableGM}.
Note however, that $\overline{G_M^b}(0)$ provides
only an upper limit.
As shown in Appendix~\ref{appIntegral},
when $M_B < M_{B'}$ one has always
${\cal I}(0) < 1$, therefore
$G_M^b(0) < \overline{G_M^b}(0)$.

The expressions (\ref{eqGMb})-(\ref{eqGCb}) show,
as mentioned already, that when we use 
the S-state approximation for the octet and decuplet wave functions,
one has only nonvanishing contributions for the
magnetic dipole form factor $G_M^\ast$.

In Table~\ref{tableGM} we also compare 
our results for $\overline{G_M^b}(0)$
with an estimate of a valence quark model (QM)~\cite{Darewych83,Leinweber93}
and the results from quenched lattice QCD~\cite{Leinweber93}
(small meson cloud effects).
Our purpose is to show that the valence quark contribution
for $G_M^\ast(0)$ is bounded and insufficient
to explain the experimental results.
Taking the $\gamma^\ast N \to \Delta$ case as an example,
our estimate, the QM result, and the lattice QCD
results give $G_M^\ast(0) \approx 2$,
while the experimental result is $3.02\pm 0.03$~\cite{PDG}.
Therefore, the valence quark contribution explains
only about 70\% of the experimental result.

For the $\gamma^* N \to \Delta$ reaction, one has also
the results of the quenched lattice simulation
from Alexandrou~{\it et al.\/}~\cite{Alexandrou08},
where the extrapolated quenched results gave,
$G_M^*(0) \simeq 2.1$ for $m_\pi=563$ MeV;
$G_M^*(0) \simeq 1.9$ for $m_\pi=490$ MeV and
$G_M^*(0) \simeq 1.8$ for $m_\pi=411$ MeV.
These results are consistent with the estimates
of the covariant spectator quark model with a
regime where the meson and baryon masses
used are those corresponding to
the lattice QCD simulations~\cite{LatticeD,Lattice}.

Recall that the results described in this section include
only the valence quark contributions.
In this case we can conclude from Table~\ref{tableJS}
that, the transitions
$\gamma^* \Sigma^+ \to \Sigma^{\ast +}$
and $\gamma^* \Xi^0 \to \Xi^{\ast 0}$,
would give the same results in the limit that
the octet ($M_B$) and decuplet ($M_{B'}$) baryon masses
are, respectively, the same for the octet and decuplet members, or
$M_\Sigma=M_\Xi$ and $M_{\Sigma^*}=M_{\Xi^*}$.
The same argument holds also for the
transitions, $\gamma^* \Sigma^- \to \Sigma^{\ast -}$
and $\gamma^* \Xi^- \to \Xi^{\ast -}$.

The above relations can also be derived
from the $U$-spin symmetry~\cite{Lipkin73}.
The $U$-spin symmetry implies
that the systems are invariant 
in the exchange of a $d$ and an $s$ quark~\cite{Keller11b,Lipkin73},
or equivalently the symmetry in the same charge multiplet.

Another interesting limit is the
exact $SU(3)$ symmetry limit,
when $f_{i\pm}(Q^2)= f_{i 0}(Q^2)\equiv f_i(Q^2)$
($f_i$ are independent of the flavors), and
all the octet baryons have a unique mass $M_B$, and all the
decuplet baryons have a unique mass $M_{B'}$.
In this limit we expect
no contributions for the reactions,
$\gamma^* \Sigma^- \to \Sigma^{\ast -}$
and $\gamma^* \Xi^- \to \Xi^{\ast -}$,
because of $j_i^S \equiv 0$,
and the same for all the other reactions,
$j_i^S= \sqrt{2} f_{i}$,
except for $\gamma^* \Sigma^0 \to \Sigma^{\ast 0}$
with $j_i^S= \sfrac{\sqrt{2}}{2} f_{i}$,
and $\gamma^* \Lambda \to \Sigma^{\ast 0}$
with $j_i^S= \sqrt{\sfrac{3}{2}} f_{i}$.
The suppression of the
contributions for the $\gamma^* \Sigma^- \to \Sigma^{\ast -}$
and $\gamma^* \Xi^- \to \Xi^{\ast -}$ reactions
compared to the others, is also obtained with the
$U$-spin symmetry~\cite{Keller11b,Lipkin73}.
Since in practice we break the $SU(3)$ symmetry
using the physical masses,
our estimates of the quark
core contributions $\overline{G_M^b}(0)$ in Table~\ref{tableGM}
can have variations of about up to 20\%, similar to the amount of
deviations in the masses.

The final result for the bare contributions
given by~(\ref{eqGMb}) should also be corrected
by a factor $\sqrt{Z_B}$ coming from
the normalization of the octet baryon wave function
due to the pion cloud effect.
This normalization is necessary to describe
the charge of the dressed baryon $B$.
As explained in Refs.~\cite{Octet,OctetFF,OctetMedium},
the quark core contribution to the electric
form factor is proportional the factor $Z_B < 1$.
Taking the proton as an example,
the pion cloud contribution for the charge is
$0.15Z_N$ (contribution from the core of $Z_N$)
in the model from Ref.~\cite{OctetMedium},
leading to $\sqrt{Z_N}=0.93$,
in order to reproduce
the total proton charge.
As for the decuplet wave functions, there
are no corrections since the model
used assumes that the pion cloud
contributions are negligible~\cite{Omega}.
We note that the effect of the octet wave function
normalization is small since $\sqrt{Z_B} \simeq 1$;
therefore it does not affect the results appreciably.
Thus, we use the simplest model, by setting $\sqrt{Z_B}=1$.
We will discuss the impact of this approximation later.

\begin{table}[t]
\begin{tabular}{l  c  r  rrr r}
\hline
\hline
    & &  $\overline{G_M^b}(0)$ &&   QM   &&  Lattice    \\
\hline
\hline
$\gamma^* p \to \Delta^+$  && 2.05 && 1.88  && 1.97(12)    \\
$\gamma^* n \to \Delta^0$  && 2.05 && 1.88 && 1.97(12)   \\
\hline 
$\gamma^* \Lambda \to \Sigma^{\ast 0}$ &&
2.02  &&  &&
\\ 
\hline
$\gamma^* \Sigma^+ \to \Sigma^{\ast +}$  &&
2.30 && 2.34 && 2.03(14) \\
$\gamma^* \Sigma^0 \to \Sigma^{\ast 0}$  && 1.07
& &  0.99 && 0.92(5)
\\
$\gamma^* \Sigma^- \to \Sigma^{\ast -}$ &&
-0.17  && -0.36 && $-$0.20(4)  \\
\hline
$\gamma^* \Xi^0 \to \Xi^{\ast 0}$ &&
2.47 && 2.72  && 2.10(8)
\\
$\gamma^* \Xi^- \to \Xi^{\ast -}$ &&
-0.19  && -0.42  && -0.202(28)
\\
\hline
\hline
\end{tabular}
\caption{
Upper limit of the magnetic dipole transition form factors
for $Q^2=0$, $\overline{G_M^b}(0)$, which are
independent of the baryon wave function parametrizations,
compared with the results of quark models~\cite{Darewych83,Leinweber93}
and lattice QCD with $m_\pi=662$ MeV~\cite{Leinweber93}. }
\label{tableGM}
\end{table}

\subsection{Pion cloud form factors}
\label{secPionCloud}

We discuss now the pion cloud contributions
for the form factors. As before, we focus on
the magnetic form factors.

\begin{figure}[t]
\vspace{.4cm}
\includegraphics[width=1.5in]{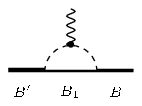}
\caption{\footnotesize
Electromagnetic interaction with the pion
(pion cloud contribution).
Note that, between the initial octet ($B$) and
the final decuplet ($B'$) baryon states, there can be
several intermediate $\pi B_1$ states.}
\label{figPionCloud}
\end{figure}

Although the pion cloud dressing 
is included at the quark level effectively
in the parametrization of the quark electromagnetic
form factors, there are processes 
involving the pion cloud that are not taken into account.
The processes which a pion is  
exchanged between the different quarks cannot be 
represented by the quark dressing due to the pion cloud.
Instead, the processes in which the pion is exchanged between  
different quarks, are regarded as the pion 
is emitted and absorbed by the {\it overall} baryon
in our model~\cite{OctetFF}, which is represented 
by the diagram in Fig.~\ref{figPionCloud}.

We assume that the dominant contribution for
the transitions comes from
the direct coupling of the photon to the pion
as depicted in Fig.~\ref{figPionCloud},
suggested by chiral perturbation theory~\cite{Arndt04}.
As a consequence the
pion cloud contributions for the
$\gamma^* B \to B'$ transitions differ
only by the quark flavor structure of the baryons, and
the kinematic effects due to the baryon masses.
In the exact $SU(3)$ limit when all the octet baryon members
have the same mass $M_B$ and also all the decuplet
baryon members have the mass $M_{B'}$,
the pion cloud contribution
will depend only on the flavor symmetry.
Namely, the flavor effect can be
determined using the $SU(3)$ meson-baryon couplings
with the $SU(6)$ symmetry mixing parameter ratio,
$\alpha \equiv D/(F+D) = 0.6$.
Thus, assuming that the loop integrals arising from
the diagram in Fig.~\ref{figPionCloud}
are only weakly dependent on the octet and decuplet baryon masses,
the pion cloud contributions for all the octet to decuplet transitions
can be estimated using the
results obtained from the $\gamma^* N \to \Delta$ transition.

In summary,
to estimate the pion cloud contributions
for the $\gamma^* B \to B'$ transitions,
we proceed as follows:
\begin{itemize}
\item
Take a parametrization established for the
pion cloud contributions for the
$\gamma^* N \to \Delta$ transition.
\item
Calculate the flavor corrections
for the $\gamma^* B \to B'$ assuming
that $M_B= M_N$ and $M_{B'}= M_\Delta$.
\end{itemize}

In the $\gamma^* N \to \Delta$ transition
the pion cloud contributions can be represented
by the phenomenological form~\cite{NDelta},
\be
G_M^\pi(Q^2) =  \lambda_\pi
\left( \frac{\Lambda_\pi^2}{\Lambda_\pi^2+ Q^2}\right)^2 (3 G_D),
\label{eqGMpi0}
\ee
where $G_D =\left(1 + \frac{Q^2}{0.71} \right)^{-2}$,
with $Q^2$ in GeV$^2$,
is the nucleon dipole form factor, $\lambda_\pi$
is a coefficient associated with
the strength of the pion cloud effect, and
$\Lambda_\pi$ a cutoff mass.
The cutoff mass $\Lambda_\pi$ controls the falloff
of the pion cloud effects.
Note that $\lambda_\pi$ gives the
relative contribution of the pion cloud to the total
magnetic form factor for small $Q^2$,
since for small $Q^2$, $G_M^* (Q^2) \approx 3 G_D$,
and $\lambda_\pi \approx \sfrac{G_M^\pi (Q^2)}{G_M^*(Q^2)}$.

To estimate the pion cloud dressing for the other
octet to decuplet transitions, it is enough to calculate
the flavor factor $f_{B B'}$ associated
with the transition $\gamma^* B \to B'$
normalized by the transitions
$\gamma^* N \to \Delta$ (or $\gamma^* p \to \Delta^+$)
as shown next.
The details are presented in Appendix~\ref{appPionCloud}.
Recalling that the strength of the pion cloud contribution
for the $\gamma^* N \to \Delta$ reaction
is given
by the value $\lambda_\pi$ at $Q^2=0$,
the corresponding strength
for the $\gamma^* B \to B'$ reaction can be obtained
with the replacement
\be
\lambda_\pi \to  f_{BB'} \, \lambda_\pi.
\ee
Thus, the pion cloud contribution
for the magnetic form factor
in the reaction $\gamma^* B \to B'$ is
\be
G_M^\pi(Q^2) =  f_{BB'} \lambda_\pi
\left( \frac{\Lambda_\pi^2}{\Lambda_\pi^2+ Q^2}\right)^2 (3 G_D).
\label{eqGMpi}
\ee
The factors $f_{BB'}$ are given in Table~\ref{tableFB}.
Note that, with the above parametrization,
we have the same $Q^2$ dependence for all
$\gamma^* B \to B'$ reactions,
which is a consequence of
assuming the $SU(3)$ symmetry for the octet
and decuplet baryon masses.

In the calculation
we use the parametrization from Refs.~\cite{NDeltaD,LatticeD}.
Explicit values are, $\lambda_\pi=0.441$ and
$\Lambda_\pi^2= 1.53$ GeV$^2$.
Although in Refs.~\cite{NDeltaD,LatticeD} there
are also higher angular momentum state contributions
(D-states) aside from the S-state
for the $\Delta$ baryon,
the effects of those states are small.
Therefore, the pion cloud parametrization
given by Eq.~(\ref{eqGMpi0})
should be a very good approximation
even when the D-states are neglected.

\begin{table}[t]
\begin{center}
\begin{tabular}{l  c   }
\hline
\hline
    & $f_{BB'}$    \\
\hline
$\gamma^* N \to \Delta$ & 1 \\ [0.05in]
$\gamma^* \Lambda \to \Sigma^{\ast 0}$ & $\sfrac{2 \sqrt{3}}{5}$ \\ [0.05in]
$\gamma^* \Sigma \to \Sigma^*$ & $\sfrac{1}{5}J_3$ \\ [0.05in]
$\gamma^* \Xi \to \Xi^*$ & $\sfrac{1}{5} \tau_3$ \\ [0.05in]
\hline
\hline
\end{tabular}
\end{center}
\caption{Coefficients $f_{BB'}$ associated with the $\gamma^* B \to B'$
transitions.
The matrices $J_3$ and $\tau_3$ are, respectively,
the third component of the isospin-1 and isospin-1/2 operators.}
\label{tableFB}
\end{table}

In summary, we
calculate the magnetic transition form
factors for the present model by,
\ba
G_M^\ast(Q^2)= G_M^b(Q^2)+ G_M^\pi(Q^2),
\label{eqGMt}
\ea
where $G_M^b$ and $G_M^\pi$ are defined
respectively by Eqs.~(\ref{eqGMb}) and (\ref{eqGMpi}).

Note that $G_M^\ast(0)$ gives the
transition magnetic moment in natural units.
To convert $G_M^\ast(0)$ into
the transition magnetic
dipole moment $\mu_{B B'}$
in nuclear magneton ($\frac{e}{2M_N}$),
we use ~\cite{Leinweber93},
\ba
\mu_{BB'} = \frac{M_N}{M_B} \sqrt{\frac{M_{B'}}{M_B}}
G_M^\ast(0) \frac{e}{2M_N}.
\ea

\section{Results}
\label{secResults}

We divide our presentation of the results and analysis
into four subsections.
We start with the discussion of the numerical results for the
transition form factors.
Next, we focus on the symmetry relations
among the different octet to decuplet transitions.
Third, we compare the results
with the available experimental information,
in particular for the reactions
aside from the $\gamma^\ast N \to \Delta$ reaction.
Finally, we discuss the overall results.

\subsection{Octet to decuplet electromagnetic transition form factors}

\begin{figure}[t]
\vspace{.4cm}
\centerline{
\mbox{
\includegraphics[width=2.8in]{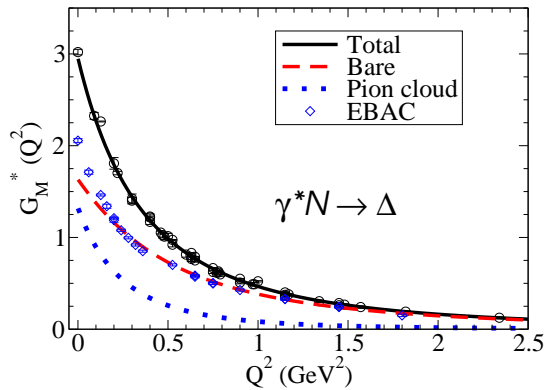}
}}
\caption{\footnotesize{
Results for the $\gamma^\ast N \to \Delta$ transition.
Data shown are for the $\gamma^\ast p \to \Delta^+$ reaction,
from DESY~\cite{Bartel68}, SLAC~\cite{Stein75},
CLAS/JLab~\cite{CLAS}
and MAID analysis~\cite{Tiator01,Drechsel07}.
Data for the large $Q^2$ region from CLAS/JLab
are not included~\cite{CLAS2}.
EBAC results are from Ref.~\cite{Diaz07a}.
}}
\label{figDelta}
\end{figure}

\begin{figure}[t]
\vspace{.4cm}
\centerline{
\mbox{
\includegraphics[width=2.8in]{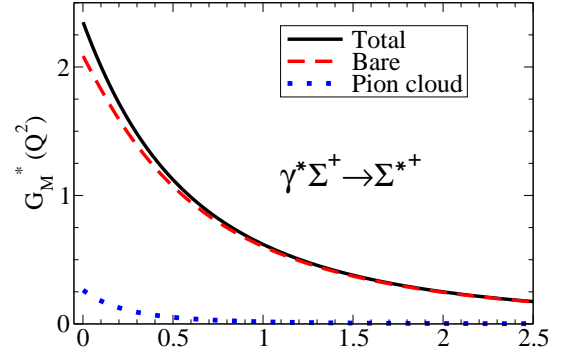}
}}
\vspace{.8cm}
\centerline{
\mbox{
\includegraphics[width=2.8in]{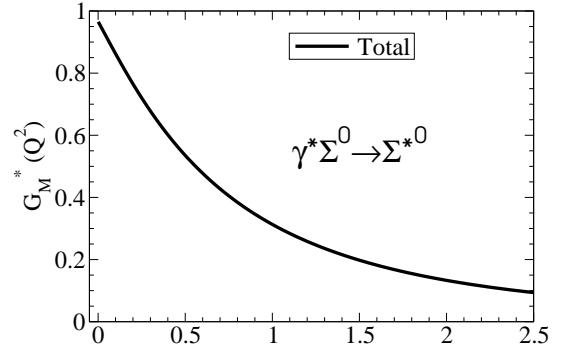}
}}
\vspace{.85cm}
\centerline{
\mbox{
\includegraphics[width=2.8in]{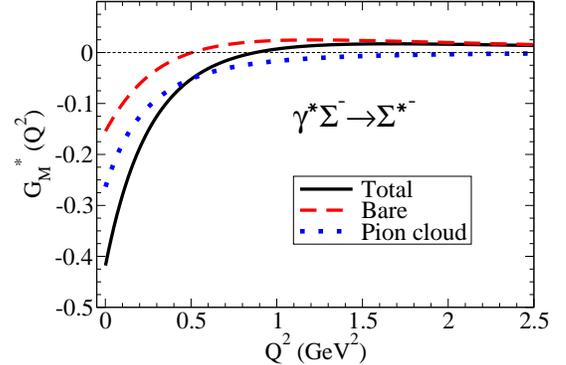}
}}
\caption{\footnotesize{
Results for the $\gamma^\ast \Sigma^{\pm,0} \to \Sigma^{*\pm,0}$ reactions.
For the $\Sigma^0$ case, the pion cloud contribution
vanishes, and the bare and
the total contributions are equal.
}}
\label{figSigma}
\end{figure}

The results of the transition form factors for the reactions,
$\gamma^\ast N \to \Delta$, $\gamma^\ast \Lambda \to \Sigma^{* 0}$,
$\gamma^\ast \Sigma \to \Sigma^{* }$, and
$\gamma^\ast \Xi \to \Xi^{* }$, are presented,
respectively in Figs.~\ref{figDelta}--\ref{figXi}.
The results for $\gamma^\ast N \to \Delta$
represent those two reactions, $\gamma^\ast p \to \Delta^+$
and $\gamma^\ast  \to \Delta^0$,
which are equal in our model.
The data are
available only for the $\gamma^\ast p \to \Delta^+$ reaction.
The calculations are based on the formulation
exposed in the previous section,
summarized by Eq.~(\ref{eqGMt}).

In Fig.~\ref{figDelta} we present
the result for the $\gamma^\ast N \to \Delta$ reaction,
including the total, and
the contributions from the bare core
(valence quark), and the pion cloud, as well as
the data for $\gamma^\ast p \to \Delta^+$
from DESY~\cite{Bartel68}, SLAC~\cite{Stein75},
CLAS/Jefferson Lab~\cite{CLAS}
and MAID analysis~\cite{Tiator01,Drechsel07}
for $Q^2 < 2.5$ GeV$^2$.
Note that, in the region $Q^2 < 2.5$ GeV$^2$,
the agreement between the model result (solid line) and
the data is excellent.
This is because the $\Delta$ wave function in the model
was calibrated previously to reproduce the data~\cite{NDeltaD,LatticeD}.
It should be mentioned however, that
the nucleon wave function used here
is different from the one used in Refs.~\cite{NDeltaD,LatticeD},
but it was obtained from the study of the octet
baryon electromagnetic form factors~\cite{OctetMedium}.
Although the pion cloud effects are
included in the treatment of the baryon systems
in Ref.~\cite{OctetMedium}
and not included in Refs.~\cite{NDeltaD,LatticeD},
both the nucleon wave functions
yield very similar results.

Figure~\ref{figDelta} also shows 
the insufficiency of the valence quark degrees
of freedom only, to reproduce the magnetic form factor.
Successful description of the reaction data
was obtained using coupled channel reaction models
(or dynamical models),
where the meson-baryon interactions are taken
into account, and the effect of the
meson cloud dressing is included.
Examples are the Sato-Lee model~\cite{SatoLee,Diaz07a},
and the Dubna-Mainz-Tapai model~\cite{Kamalov}.
Also in these cases the pion cloud is about 30-45\% of the total. 
See Refs.~\cite{Pascalutsa07,Burkert04a} for a review.

Included also in Fig.~\ref{figDelta} is the
estimate of the quark core contributions
from the EBAC group based on the
Sato-Lee model~\cite{Diaz07a}.
The results are obtained using
the Sato-Lee model,
when the pion cloud contributions are removed.
The good agreement between our bare result (dashed line)
and the EBAC result,
apart from the small deviation in the region
$Q^2<0.2$ GeV$^2$, is an indication that
our parametrization~(\ref{eqGMpi0})
gives a good representation of the pion cloud effects.

The results for $G_M^\ast$,
together with the bare, and the pion cloud contributions
for $Q^2=0$, are presented in Table~\ref{tableGM0}.
The comparison of the bare, $G_M^b(0)$, with
the upper limit, $\overline{G_M^b} (0)$,
in Table~\ref{tableGM}, allows us to conclude that
the valence quark contribution in the model gives only
about 80-90\% of the maximum value.

In Fig.~\ref{figLambda} one can see the dominance
of the valence quark (bare) contribution
in the $\gamma^\ast \Lambda \to \Sigma^{* 0}$ reaction.
This feature is expected based on the estimate
of the pion cloud contribution:
about 0.92 at $Q^2=0$, smaller than
the one for the $\gamma^* N \to \Delta$
transition of 1.32 as shown in Table~\ref{tableGM}.

As for the $\gamma^* \Sigma^{\pm,0} \to \Sigma^{*\pm,0}$ reactions,
one can observe in Fig.~\ref{figSigma}
different trends by the $\Sigma$ charges.
For the reaction with the $\Sigma^+$, the result is comparable with
that of the $\gamma^* N \to \Delta$,
while one has a smaller magnitude of about 50\%
for the reaction with the $\Sigma^0$, and an even smaller magnitude
for the reaction with the $\Sigma^-$.
In these reactions the magnitude of the pion cloud contributions
is small: 0.26 at $Q^2=0$ (about 20\%
of the $\gamma^* N \to \Delta$ reaction)
for the reactions with the $\Sigma^{\pm}$,
and vanishes for the reaction with the $\Sigma^0$.

The results for the $\gamma^* \Xi^{0,-} \to \Xi^{*0,-}$ reactions
are presented in Fig.~\ref{figXi}. They are similar
to the results described for the
$\gamma^* \Sigma^+ \to \Sigma^{*+}$
and $\gamma^* \Sigma^- \to \Sigma^{*-}$, respectively
for the reactions with the $\Xi^0$ and
$\Xi^-$ in the initial states.

In Figs.~\ref{figDelta}--\ref{figXi},
we can observe the fast falloff of the
pion cloud contributions, and the dominance
of the valence quark contributions with increasing $Q^2$.
For a very large $Q^2$, one has $G_M^\ast \propto 1/Q^4$
according to Eq.~(\ref{eqGMb}), in agreement with
pQCD estimates~\cite{NDelta,Carlson}.
The pion cloud contributions given by Eq.~(\ref{eqGMpi}),
vary as $G_M^\pi \propto 1/Q^8$.

We now comment on the effects due to the
baryon wave function normalization.
As mentioned already, only the octet baryon wave functions
are subject to be modified in the present treatment,
by the factor $\sqrt{Z_B}$ in the
valence quark contributions.
The effect of the normalization is in general small,
since $\sqrt{Z_B} \simeq 1$
($\sqrt{Z_\Lambda}=0.965$,
$\sqrt{Z_\Sigma}=0.958$ and  $\sqrt{Z_\Sigma}=0.997$),
except for the core contribution for
the $\gamma^\ast N \to \Delta$ transition
with a 7\% correction ($\sqrt{Z_N}= 0.931$).
In this case, however, the effect in the total magnitude of the
form factor is about 3\%,
because the correction affects only the bare contribution
and the pion cloud contribution is significant (44\%).
Thus, we conclude that the corrections
due to the normalization of the baryon wave functions
are small (order of a few percent), and
can be neglected in a first approximation.

\subsection{Symmetry between different transitions}

Roughly, we can classify the results for
the $\gamma^* B \to B'$ transition form
factors according to the magnitudes of magnetic dipole
form factor $G^*_M$:
\ba
& &
\mbox{large}: \gamma^* N \to \Delta,
\gamma^* \Lambda \to \Sigma^{*0}, \nonumber \\
& &
\hspace{1cm} \gamma^* \Sigma^+ \to \Sigma^{*+},
\gamma^* \Xi^0 \to \Xi^{*0},
\nonumber \\
& &
\mbox{moderate}: \gamma^* \Sigma^0 \to \Sigma^{*0},
\nonumber \\
& &
\mbox{small}: \gamma^* \Sigma^- \to \Sigma^{*-},
  \gamma^* \Xi^- \to \Xi^{*-}.
\nonumber
\ea
This classification has an implication for the magnitudes
of the decay widths as we will see in the next section.

\begin{figure}[t]
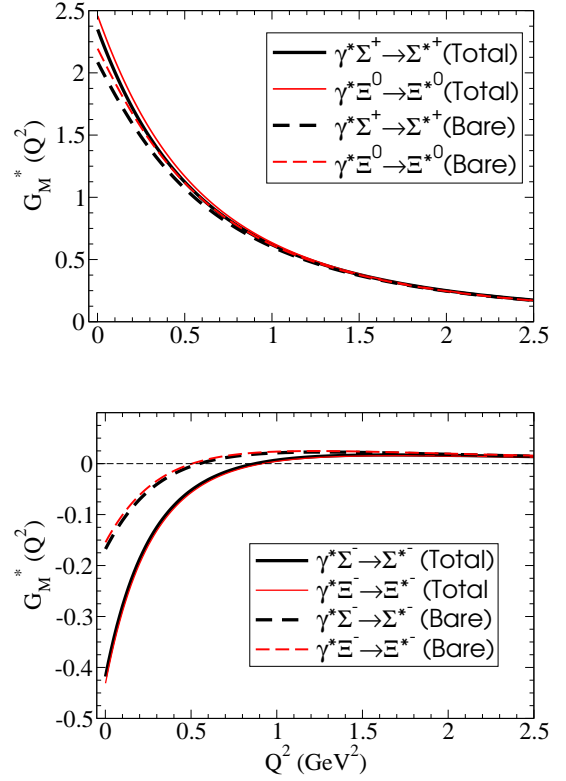

\vspace{.4cm}
\centerline{
\mbox{
\includegraphics[width=2.8in]{SigmaXi+} }}
\vspace{.85cm}
 \centerline{
\mbox{
\includegraphics[width=2.8in]{SigmaXi-}
}}
\caption{\footnotesize{
Comparison between the
$\gamma^\ast \Sigma^+ \to \Sigma^{\ast +}$ and
$\gamma^* \Xi^{0} \to \Xi^{*0}$ reactions (top)
and between
the $\gamma^\ast \Sigma^- \to \Sigma^{\ast -}$ and
$\gamma^* \Xi^{0} \to \Xi^{*0}$ reactions (bottom).
}}
\label{figSigmaXi}
\end{figure}

The observed magnitudes for $G^*_M$ mainly reflect
the dominant valence quark structure,
although modified by the effect of the pion cloud.
As mentioned in Sec.~\ref{secValence}
based on Table~\ref{tableJS}, except for
the deviations due to the mass differences,
we can expect similar results for the
$\gamma^\ast \Sigma^+ \to \Sigma^{* +}$
and $\gamma^\ast \Xi^0 \to \Xi^{* 0}$ transitions.
The same holds for the reactions
$\gamma^\ast \Sigma^- \to \Sigma^{* -}$
and $\gamma^\ast \Xi^- \to \Xi^{* -}$.
We compare the results for these reactions
directly in Fig.~\ref{figSigmaXi}.

Note in Fig.~\ref{figSigmaXi}, the closeness between
the results for the two reactions both for the bare (dashed lines)
and the total (solid lines).
These results are the consequences of the following two effects:
similarity in the valence quark structure,
and identical contribution from the pion cloud
contributions (see Table~\ref{tableFB}).
Concerning the valence quark contributions,
the similarity in the results of the two reactions
is a combination of the identical transition
current coefficients ($j_i^S$)
and the kinematics.
In fact, although the mass configurations
are different for the $\gamma^* \Sigma \to \Sigma^*$
and  $\gamma^* \Xi \to \Xi^*$ reactions,
the transition three-momentum $|{\bf q}|$
at $Q^2=0$ in the baryon $B'$ rest frame,
are almost the same,
0.18 GeV and 0.20 GeV respectively.

The difference in magnitude
between the two sets,
($\gamma^\ast \Sigma^+ \to \Sigma^{* +}$,
$\gamma^\ast \Xi^0 \to \Xi^{* 0}$)
and
($\gamma^\ast \Sigma^- \to \Sigma^{* -}$,
$\gamma^\ast \Xi^- \to \Xi^{* -}$)
in our model, is a consequence of the approximate
$SU(3)$ symmetry.
Furthermore, as commented in Sec.~\ref{secValence},
a model with the exact $SU(3)$ symmetry limit would give no
contribution for the last two reactions.
In contrast, the small violation of
the symmetry, in particular in the $SU(2)$ sector
due to the asymmetry between the isoscalar and isovector
quark form factors $f_\pm(Q^2)$,
is the reason why the present model is
successful in the description
of the neutron electric form factor~\cite{OctetFF,OctetMedium,Nucleon}.
In other approaches the small magnitude
of the $G^*_M$ results for the
$\gamma^\ast \Sigma^- \to \Sigma^{* -}$ and
$\gamma^\ast \Xi^- \to \Xi^{* -}$ reactions,
can be a consequence of $U$-spin symmetry~\cite{Lipkin73}.

\begin{table}[t]
\begin{center}
\begin{tabular}{l  r r  r c }
\hline
\hline
     &  \sp\sp$G_M^b(0)$ \sm\sm  & \sp\sp $G_M^\pi(0)$ \sm\sm & \sp\sp $G_M^*(0)$\sm\sm &  $|G_M^*(0)|_{\rm exp}$\\
\hline
$\gamma^* p \to \Delta^+$  & 1.63 &  1.32  &  2.95 & \sp\sp$3.04\pm0.11$ \cite{PDG}\\
$\gamma^* n \to \Delta^0$  & 1.63 &  1.32  &  2.95 & \sp\sp$3.04\pm0.11$ \cite{PDG}\\
\hline
$\gamma^* \Lambda \to \Sigma^{\ast 0}$ & 1.68 &  0.92 & 2.60   & \sp\sp$3.35\pm0.57$ \cite{PDG}\\
\hline
$\gamma^* \Sigma^+ \to \Sigma^{\ast +}$ & 2.09 & 0.26  & 2.35   & \sp\sp$4.10\pm0.57$
\cite{Keller11a}\\
$\gamma^* \Sigma^0 \to \Sigma^{\ast 0}$ & 0.97 & 0.00  & 0.97   &  \\
$\gamma^* \Sigma^- \to \Sigma^{\ast -}$ & $-0.15$ & $-0.26$  &$-0.42$   &  $< 0.8$
\cite{Molchanov04} \\
\hline
$\gamma^* \Xi^0 \to \Xi^{\ast 0}$ & 2.19  & 0.26  & 2.46  &   \\
$\gamma^* \Xi^- \to \Xi^{\ast -}$ & $-0.17$ & $-0.26$  & $-0.43$   & \\
\hline
\hline
\end{tabular}
\end{center}
\caption{Results for $G_M^*(0)$. Values for
$|G_M^*(0)|_{\rm exp}$ are estimated by Eq.~(\ref{eqGamma})
using the experimental values of $\Gamma_{B' \to \gamma B}$.}
\label{tableGM0}
\end{table}

\begin{figure}[t]
\vspace{.4cm}
\centerline{
\mbox{
\includegraphics[width=2.8in]{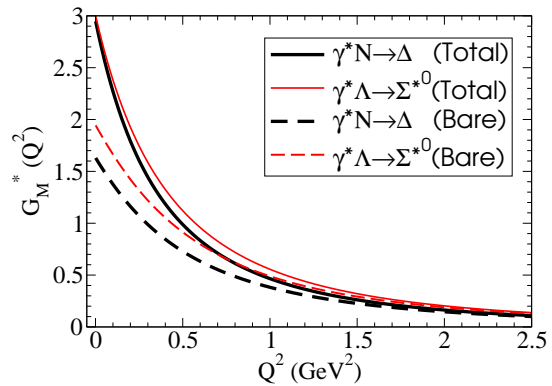}
}}
\caption{\footnotesize{
Comparison of the reactions $\gamma^* N \to \Delta$
and $\gamma^* \Lambda \to \Sigma^{* 0}$.
The results for $\gamma^* \Lambda \to \Sigma^{* 0}$
are multiplied by the factor $\sqrt{\frac{4}{3}}$.
}}
\label{figDeltaLambda}
\end{figure}

We can also study the relation between
the transitions $\gamma^* N \to \Delta$
and $\gamma^\ast \Lambda \to \Sigma^{*0}$
based on the similarity suggested by
the valence quark structure given in
Table~\ref{tableJS}.
From Table~\ref{tableJS}, we may conclude
that the transition form factors
between the $\gamma^\ast \Lambda \to \Sigma^{*0}$
and  $\gamma^* N \to \Delta$ reactions differ
by a factor $\sqrt{\frac{3}{4}}$,
if only the valence quark contributions
are considered.
We examine this in Fig.~\ref{figDeltaLambda},
by comparing the form factor of $\gamma^* N \to \Delta$
to that of $\gamma^\ast \Lambda \to \Sigma^{*0}$
multiplied by $\sqrt{\frac{4}{3}}$.
However, the results must be interpreted with care.
Focusing on the final results (total, solid lines),
the similarity between the results
for the two reactions is an accidental
combination of a large pion cloud effect
and a smaller core contribution
for the $\gamma^* N \to \Delta$ reaction,
and the opposite,
a smaller pion cloud effect and a larger core contribution
for the $\gamma^\ast \Lambda \to \Sigma^{*0}$ reaction.
The symmetry properties should be better
observed in the bare contributions (dashed lines).
In fact, the two dashed lines have
a similar shape, but differ in magnitudes by
about 20\% near $Q^2=0$.
This is a consequence of
the differences in the masses and radial wave functions.

\begin{figure}[t]
\vspace{.4cm}
\centerline{
\mbox{
\includegraphics[width=2.8in]{Lambda1a.eps}
}}
\caption{\footnotesize{
Results for the $\gamma^\ast \Lambda \to \Sigma^{* 0}$ transition.
}}
\label{figLambda}
\end{figure}

Then, we conclude that
the closeness between the total results
for the $\gamma^* N \to \Delta$ and
$\gamma^\ast \Lambda \to \Sigma^{*0}$ reactions,
also predicted by the $U$-spin symmetry,
is accidental, since the pion cloud contributions
should break the symmetry appreciably.
In fact, for the $\gamma^* N \to \Delta$ reaction,
the pion cloud contribution is
80\% of the quark core contribution,
while in the $\gamma^\ast \Lambda \to \Sigma^{*0}$ reaction,
the pion contribution is 55\%.
Note that, the $U$-spin symmetry takes into account only
the valence quark contributions of the baryons.
If it is applied also for the meson cloud contributions,
one must assume the same proportionality between
the meson cloud and valence quark contributions.

\subsection{Decay widths}

We now discuss the results for the
$B' \to \gamma B$ decay widths,
which is closely connected with $|G_M^\ast(0)|$, as we show next.
Therefore, the discussion about the decay widths
is nearly equivalent to the discussion of the
magnitudes, $|G_M^\ast(0)|$.
Note, however, that only the decay widths for the
reactions, $\gamma^\ast N \to \Delta$,
$\gamma^\ast \Lambda \to \Sigma^0$, and
$\gamma^\ast \Sigma^+ \to \Sigma^+$
are experimentally determined.

Assuming the $G_M^*$ dominance
($G_E^*, G_C^* \simeq 0$),
we can calculate the decay
width $\Gamma_{B' \to \gamma B}$ by~\cite{Leinweber93,Wolf90,DeltaTL},
\ba
\Gamma_{B' \to \gamma B}=
\frac{\alpha}{16} \frac{(M_{B'}^2-M_B^2)^3}{M_{B'}^3 M_B^2} |G_M^*(0)|^2,
\label{eqGamma}
\ea
where $\alpha = \sfrac{e^2}{4\pi}\simeq \sfrac{1}{137}$
is the electromagnetic fine structure constant.

The assumption of the $G_M^*$ dominance for $Q^2=0$
is justified when $|G_E^\ast(0)|$ is
small enough, e.g., it is an order of a few percent of $|G_M^*(0)|$.
Then, since the correction to the term $|G_M^\ast(0)|^2$
in Eq.~(\ref{eqGamma}) enters as
$|G_M^\ast(0)|^2+ 3 |G_E^\ast(0)|^2$,
we may neglect the $G_E^\ast(0)$
with an accuracy of about 1\%.
This is indeed supported by the different estimates for
$G_E^\ast(0)$~\cite{Abada96,Butler93b,Haberichter97,Wagner98,Leinweber93}.

Estimates of $G_M^*(0)$ based on Eq.~(\ref{eqGamma})
are presented in Table~\ref{tableGM0},
together with our predictions for $G_M^*(0)$.
Notice in particular,
the result for the $\gamma^\ast N \to \Delta$ reaction,
$G_M^\ast(0)=3.04\pm 0.11$, is very close to
the experimental value of
$G_M^*(0)=3.02\pm 0.03$~\cite{Tiator01}.
In Table~\ref{tableGM0} we can see that
our results for $\gamma^\ast \Lambda \to \Sigma^{* 0}$
and $\gamma^\ast \Sigma^+ \to \Sigma^{* +}$
underestimate the values of $|G_M^\ast(0)|$,
determined from the data.
In fact our results give only 78\% and
57\% respectively, compared with the
corresponding experimental central values
(underestimates of 1.3 and 3.1 standard deviations respectively).
We will discuss the impact of these results
in more detail later.

In the S-state approximation with $G_E^*=0$,
we can also calculate the helicity amplitudes $A_{3/2}$
and $A_{1/2}$ in terms of $G_M^*$
using the relations, $A_{3/2}= -\sqrt{3}{F} G_M^*$
and  $A_{1/2}= -F G_M^*$, where the factor
$F$ is a given function of $Q^2$~\cite{Delta1600,Capstick95}.
For $Q^2=0$, the factor $F$ is given by
\mbox{$F= \frac{e}{4 M_B} \sqrt{\frac{M_{B'}^2-M_B^2}{2M_B}}$.}
In Table~\ref{tableGamma} we present
the results for the helicity amplitudes
$A_{3/2}$ and $A_{1/2}$ for $Q^2=0$,
calculated in the approximation $G_E^*=0$.
Finally, we also present 
our predictions for the decay widths
calculated by Eq.~(\ref{eqGamma}).
The results are compared with the available
experimental results for $\Gamma_{B' \to \gamma B}$ ($\Gamma_{\rm exp}$).

\begin{figure}[t]
\vspace{.4cm}
\centerline{
\mbox{
\includegraphics[width=2.8in]{Xi01a.eps}
}}
\vspace{.8cm}
\centerline{
\mbox{
\includegraphics[width=2.8in]{Xi-1a.eps}
}}
\caption{\footnotesize{
Results for the $\gamma^\ast \Xi^{0,-} \to \Xi^{*0,-}$ reactions.
}}
\label{figXi}
\end{figure}

Our results for the decay widths in Table~\ref{tableGamma}
are comparable with most of the predictions
presented in the  literature~\cite{Darewych83,Kaxiras85,Warns,Sahoo95,Wagner98,Sharma10,Schat96,Abada96,Haberichter97,Wang09,Lebed11,Butler93a,Tiator01}.
The exception is the result
for the $\Delta \to \gamma N$ reaction,
where most of the models underestimate
the experimental data by more than
200 keV~\cite{Darewych83,Wagner98,Sharma10,Sahoo95,Abada96,Haberichter97,Bijker00},
except for the Heavy Baryon Chiral Perturbation Theory
(HB$\chi$PT)~\cite{Butler93a},
large $N_c$ limit~\cite{Lebed11} and
QCD sum rules~\cite{Wang09}.

Estimates for the $\Sigma^{*0} \to \gamma \Lambda$
decay width are in the range 150--300 keV,
for a large variety of quark models,
algebraic models of hadron structure,
Skyrme, and soliton models~\cite{Darewych83,Kaxiras85,Warns,Sahoo95,Wagner98,Sharma10,Schat96,Abada96,Haberichter97}.
Only HB$\chi$PT has a window 252--540 keV~\cite{Butler93a},
while the large $N_c$ limit predicts $336\pm81$ keV~\cite{Lebed11}
and may overestimate the result of 300 keV,
as well as the QCD sum rules with 409 keV~\cite{Wang09}.
Our result, 284 keV,
underestimates the experimental value from Ref.~\cite{PDG} by
1.2 standard  deviations, and also
that from Refs.~\cite{Keller11a,Keller11b} by
1.6 standard deviations.

As for the $\Sigma^{*+} \to \gamma \Sigma^+$ decay width,
most of the predictions are in the range 50--110 keV,
with the following exceptions:
QCD sum rules (150 keV)~\cite{Wang09},
an algebraic model of hadron structure
(141 keV)~\cite{Bijker00},
HB$\chi$PT (70--220 keV)~\cite{Butler93a},
and large $N_c$ limit ($140\pm36$ keV)~\cite{Lebed11}.
Overall, these estimates are
considerably smaller than
the experimental result of $250\pm 70$ keV~\cite{Keller11a},
except for HB$\chi$PT~\cite{Butler93a}.
Our estimate, 82 keV,
underestimates the data more than 2.4 standard deviations.

As for the remaining reactions,
no experimental data are available, and
the decay widths we have obtained, are comparable with
those calculated by the several theoretical models.
In particular, the $\Xi^{*0} \to \gamma \Xi^0$
decay width is close to the $\Sigma^{*+} \to \gamma \Sigma^+$
result (82 keV versus 101 keV, in our case);
the $\Sigma^{*0} \to \gamma \Sigma^0$ decay width (14 keV)
is about an order of magnitude
smaller than that for $\Sigma^{*+} \to \gamma \Sigma^+$;
and the results for $\Sigma^{*-} \to \gamma \Sigma^-$ and
$\Xi^{*-} \to \gamma \Xi^-$ are reduced to a few keV
(2.6 keV and 3.6 keV, in our case).

Concerning the
$\gamma^\ast \Lambda \to \Sigma^{* 0}$ and
$\gamma^\ast \Sigma^+ \to \Sigma^{* +}$
decay widths, they can also be compared with the
estimates made based on the $U$-spin symmetry.
The $U$-spin symmetry relates
the $\gamma^\ast \Lambda \to \Sigma^{* 0}$ and
$\gamma^\ast \Sigma^+ \to \Sigma^{* +}$ reactions
with the $\gamma^\ast N \to \Delta$ reaction.
One can then make predictions for the
$\gamma^\ast \Lambda \to \Sigma^{* 0}$ and
$\gamma^\ast \Sigma^+ \to \Sigma^{* +}$ reactions
using the experimental results for $\gamma^\ast N \to \Delta$.
Assuming that the $U$-spin symmetry
holds for the $|G_M^\ast(0)|$,
we obtain using Eq.~(\ref{eqGamma}),
$292\pm 27$ keV
for $\gamma^\ast \Lambda \to \Sigma^{* 0}$,
and $138\pm 13$ keV for
$\gamma^\ast \Sigma^+ \to \Sigma^{* +}$.
Note the closeness of the result for
$\gamma^\ast \Lambda \to \Sigma^{* 0}$ with
our result.
The difference is 0.8 standard deviations.
As for the  $\gamma^\ast \Sigma^+ \to \Sigma^{* +}$ reaction,
our prediction, 82 keV, underestimates the
$U$-spin symmetry estimate by 4.3 standard deviations
as a consequence of the small errorbar,
although the result deviates only about 41\% from the central value.

A comment on the $U$-spin symmetry
estimates made for the same reactions in Ref.~\cite{Keller11b}
is in order.
Those estimated values are in close agreement
with the experimental results.
We note, however that the estimates in Ref.~\cite{Keller11b}
were based on the $U$-spin symmetry
between the helicity amplitudes and not
between the form factors $G_M^\ast(0)$ as discussed previously
in our case.
The difference in both estimates are the mass
factors included in the coefficient,
$F= \frac{e}{4 M_B} \sqrt{ \frac{M_{B'} |{\bf q}|}{M_B}}$,
which transforms form factors into helicity amplitudes.

\subsection{Discussion}

The interpretation of the gap between
our results for the decay widths
and the experimental ones
can be more easily made using $G_M^\ast(0)$,
assuming that $G_M^\ast$ is the dominant form factor.
As discussed already, the $G_M^\ast$ dominance is
indeed a good approximation.

Based on the upper limits for the bare results $G_M^b(0)$
given in Table~\ref{tableGM} represented as $\overline{G^b_M}(0)$,
we can conclude that the core contribution
is at most 2.0 for $\gamma^* \Lambda \to \Sigma^{* 0}$,
and 2.3 for $\gamma^* \Sigma^+ \to \Sigma^{* +}$.
These limits are the consequence of the normalization
of the baryon wave functions, and cannot be exceeded
if only the valence quark degrees of freedom are considered.
In the following we
assume that the experimental
sign of $G_M^\ast(0)$ is the same as $G_M^b(0)$.
In Table~\ref{tableGM0} comparing our results
with the {\it experimental estimates}
of 3.35 and 4.10, respectively, for
$\gamma^* \Lambda \to \Sigma^{* 0}$
and $\gamma^* \Sigma^+ \to \Sigma^{* +}$,
one can conclude that about $1.35\pm0.57$
for the former, and $1.80 \pm 0.57$ for the latter,
should be a consequence of other
effects than the valence quarks,
such as the pion (or meson) cloud effects.
However, our present estimate of the pion cloud effects
is very small for the $\gamma^* \Sigma^+ \to \Sigma^{* +}$
reaction (only 0.26), leading to a noticeable underestimate
of the experimental value of $4.1\pm0.6$.
The necessary amount of the pion cloud would be
then $1.8\pm0.6$, where the lower limit
1.2, is roughly the same amount of the pion cloud in the
$\gamma^\ast N \to \Delta$ reaction.
This minimum amount necessary is much larger than
the 0.26 of our present estimate.
Furthermore, notice that the above estimate
is made using the upper limit value
for $G_M^b(0)$, which is independent of the radial wave functions.
If we use the values for $G_M^b(0)$
given in Table~\ref{tableGM0}, the
amount of the {\it missing} meson cloud contribution should
be even larger ($2.0\pm0.6$ for the
$\gamma^\ast \Sigma^+ \to \Sigma^{*+}$ case).

Then, we conclude that even larger pion or other
meson cloud contributions are necessary to explain
the  $\gamma^\ast \Sigma^+ \to \Sigma^{*+}$ data,
in particular the decay width.
We emphasize that this conclusion
is not only restricted to our model, but
can also be inferred from a large variety of theoretical models.
As mentioned already, typical predictions
for the $\Sigma^{*+} \to \gamma \Sigma^+$ decay width
are in the 50--110 keV range,
where the more optimistic estimates differ
from the experimental value by
2 standard deviations.
Similarly, the $U$-spin symmetry estimate differs
by 1.6 standard deviations.

Therefore, the study of the
$\gamma^\ast B \to B'$ reactions requires
more elaborated investigations.
A possible effect to rescue the shortage
of the present result for
$\Sigma^{\ast +} \to \gamma \Sigma^+$ decay width,
and not yet included in our model,
is the kaon cloud contributions.
Although the contributions may be negligible
for the $\gamma^\ast N \to \Delta$
reaction, the kaon cloud effects are expected to be larger
in
the $\gamma^\ast \Xi^{0,-} \to \Xi^{\ast 0,-}$ reactions,
and can also be important
for
the $\gamma^\ast \Sigma^{0,\pm}
\to \Sigma^{\ast 0,\pm}$ reactions,
because of the strangeness.

A simple estimate based on the exact $SU(3)$ symmetry
predicts that the kaon cloud cloud contribution
is 1/6 of the pion cloud contribution
for the $\gamma^\ast N \to \Delta$ reaction.
The same estimate for the
$\gamma^\ast \Sigma^+ \to \Sigma^{* +}$ reaction
gives a kaon cloud contribution
of five times larger than that of the pion cloud in this limit,
which increases the total meson cloud contributions
(pion plus kaon) to the same amount of the
$\gamma^\ast N \to \Delta$ transition.
This enhancement of the meson cloud contributions
would increase our estimate for the
$\Sigma^{* +} \to \gamma \Sigma^+$ decay width
for a value compatible with the experimental result.
Note, however, the $SU(3)$ symmetry is
broken in nature; namely, the kaon
is heavier than the pion,
and the kaon cloud contribution
should be smaller than the estimate based on
the $SU(3)$ symmetry.
Nevertheless, it is worth noting that
the kaon cloud contribution should
increase our result
and lead to a better agreement with the data.
The kaon cloud correction affects also
the $\gamma^\ast \Xi^{0,-} \to \Xi^{\ast 0,-}$ reactions,
and in a smaller amount the $\gamma^\ast N \to \Delta$ reaction.

A more realistic estimate of the kaon cloud
contributions for the electromagnetic
transition form factors,
including explicitly the dependence
on the masses of the kaon, the octet
and decuplet baryons,
is a very promising
topic of investigation for the future.
Such a study may help to explain
the decuplet decay widths but
it is beyond the scope of the present work.
In this exploratory study, we have focused
on the valence quark and the pion cloud contributions.

In summary, the present study suggests that
the meson cloud effects, besides the pion cloud,
are important in the $\gamma^\ast B \to B'$
reactions, in particular for
those involving the $\Sigma^\ast$ in the final states.
As pointed out already in Refs.~\cite{Keller11a,Keller11b},
meson cloud effects may be indispensable to explain the data.
In fact, lattice QCD simulations,
quark models, and others generally underestimate the
magnitude of the form factors extracted from the data.
Even the models with the pion
cloud effects~\cite{Sharma10,Schat96,Abada96,Haberichter97}
fail to reproduce the magnitude for the
$\gamma^* \Lambda \to \Sigma^{* 0}$
and  $\gamma^* \Sigma^+ \to \Sigma^{* +}$ form factors.
The estimates from HB$\chi$PT, where the kaon cloud
was taken into account, support also the relevance
of the kaon cloud effects~\cite{Butler93a}.
The quantitative estimates for the decay widths
from HB$\chi$PT
(252--540 keV for $\gamma^* \Lambda \to \Sigma^{* 0}$
and 70--220 keV for $\gamma^* \Sigma^+ \to \Sigma^{* +}$),
are, however, too broad to draw more definite conclusions.
More accurate estimates of the kaon cloud
effects may help to explain the gap existing between
the predictions and the data.

In order to clarify and improve the present situation,
new experimental determinations of the decuplet to
octet radiative decay widths would be very useful.
Of critical importance is to confirm
(or deny) the result for the
$\Sigma^{*+} \to \gamma \Sigma^+$ decay width.
The determination of the other decuplet baryon radiative
decay widths can also be important.
For instance, the determination of the $\Sigma^{*0} \to \gamma \Sigma^0$
decay width would be an excellent test for theoretical models,
in particular to clarify the role of the meson cloud,
since the valence quark contribution is expected to be very small.
Another interesting case would be
the determination of the $\Xi^{*0} \to \gamma \Xi^0$ decay width,
because it is expected to be close to that of
$\Sigma^{*+} \to \gamma \Sigma^+$
in our model, and also according to the $U$-spin symmetry.

\section{Summary and conclusions}
\label{secConclusions}

In this work we have studied the octet to decuplet baryon electromagnetic
transitions using the covariant spectator quark model,
and predicted the magnetic dipole form factors for
the reactions with strange baryons.
In the present study we have adopted well established parametrizations
for the octet and decuplet baryon wave functions
developed in the previous works.
Our estimates of the valence quark contributions
for the transition form factors are based on the assumption
that the quark-diquark S-state
is the dominant configuration in the baryon systems.
Our results are consistent with lattice QCD simulations
and those of other quark models.
Based on the $SU(3)$ symmetry
for the meson-baryon couplings, we have extended
the calculation of the pion cloud contributions
for the $\gamma^\ast N \to \Delta$ reaction
to the remaining $\gamma^\ast B \to B'$ reactions
with strange baryons.

It would also be very interesting to go
beyond the S-state approximation and estimate
the quadrupole form factors. 
However, except for the $\Delta$ case,
one has no reliable parametrization at the moment 
for the small D-state components 
in the decuplet baryon wave functions, which yield 
the contributions for those form factors~\cite{NDeltaD,LatticeD,Omega}.
Nevertheless, the contributions from such small components  
are expected to be only of the order of a few percent 
compared to the magnetic dipole form factor.

It is shown that the covariant spectator quark model
is very useful to estimate the valence quark contributions
for the $\gamma^\ast B \to B'$ transition form factors,
since, in particular, it provides an upper limit of
the valence quark contributions independent of the details
of the baryon radial wave functions,
that can be used to infer the magnitude
of other contributions besides
the valence quark contributions.
In particular, the estimate of the valence quark contribution
for the $\gamma^\ast N \to \Delta$ reaction
is very important to understand why the pion cloud,
or meson cloud in general,
is of fundamental importance to obtain
a consistent description of the experimental data.

When compared with the available experimental data
(including the $\gamma^* \Lambda \to \Sigma^{* 0}$ and
$\gamma^* \Sigma^+ \to \Sigma^{* +}$ reactions),
we have found that the valence quark plus pion cloud
contributions are insufficient to explain the data.
This shortcoming is particularly evident
for the $\gamma^* \Sigma^+ \to \Sigma^{* +}$ reaction.
Namely, our result underestimates the experimental
radiative decay width by 2.4 standard deviations.
Since the effect of the valence quark core is
bounded by an upper limit, we interpret
the underestimates of the present study
as a consequence of the smallness
of the meson cloud contributions in the model,
where we have included explicitly only
the effect of the lightest meson,
the pion in this exploratory study
where it is generally believed to be dominant.

Our results strongly suggest the potential importance
of including clouds of mesons heavier than the pion
in the $\gamma^\ast B \to B'$ transitions with strange baryons,
especially the kaon cloud.
A simple estimate based on the $SU(3)$ symmetry
using the same mass and couplings for the kaon and pion,
indicates that the meson cloud contributions,
pion plus kaon clouds, are expected to increase the magnitude
for the $\Sigma^{*+} \to \gamma \Sigma^+$ form factors,
and improve the present result
towards the experimental one.
As the $SU(3)$ symmetry is broken in practice,
we conclude that a more elaborate
and consistent study for the meson cloud dressing
is necessary in order to understand better
the $\gamma^\ast B \to B'$ data.

Finally, we emphasize again that,
more experimental data for the octet to decuplet baryon transitions,
$\gamma^\ast B \to B'$, are desired to clarify
the present situation,
and shed light on the reaction mechanisms.
%

\begin{table}[t]
\begin{center}
\begin{tabular}{l  r r  r c c}
\hline
\hline
    & \sp\sp$G_M^*(0)$ \sm\sm & \sp\sp $A_{3/2}(0)$ \sm\sm\sm & \sp\sp$A_{1/2}(0)$
\sm\sm\sm& \sp\sp$\Gamma_{B' \to \gamma B}$ & $\Gamma_{\rm exp}$\\
\hline
$\gamma^* p \to \Delta^+$ & 2.95 & $-240$ &  $-139$  & 620 & $660\pm 47$ \cite{PDG}\\
$\gamma^* n \to \Delta^0$ & 2.95 & $-240$ &  $-139$  & 620 &$660\pm 47$ \cite{PDG}\\
\hline
$\gamma^* \Lambda \to \Sigma^{\ast 0}$ & 2.60 & $-168$  & $-97$ & 284 & $470\pm160$ \cite{PDG}\\
                                    &  &   &    &  & $445\pm102$
\cite{Keller11a,Keller11b}\\
\hline
$\gamma^* \Sigma^+ \to \Sigma^{\ast +}$ & 2.35 & $-118$  & $-68$   & 82 & $250\pm70$
\cite{Keller11a}\\
$\gamma^* \Sigma^0 \to \Sigma^{\ast 0}$ & $0.97$ & $-48$  & $-28$   & 14 & \\
$\gamma^* \Sigma^- \to \Sigma^{\ast -}$ & $-0.42$ & 21  & 12  & 2.6 & $<9.5$
\cite{Molchanov04} \\
\hline
$\gamma^* \Xi^0 \to \Xi^{\ast 0}$ & 2.46 & $-118$  & $-68$  & 101 &\\
$\gamma^* \Xi^- \to \Xi^{\ast -}$ & $-0.43$ & 21  & 12  & 3.1 &\\
\hline
\hline
\end{tabular}
\end{center}
\caption{Results for $G_M^*(0)$, helicity amplitudes
$A_{3/2}(0)$ and $A_{1/2}(0)$ in $10^{-3}$ GeV$^{-1/2}$,
and decay widths $\Gamma_{B' \to \gamma B}$ in keV.   }
\label{tableGamma}
\end{table}

\vspace{0.2cm}
\noindent
{\bf Acknowledgments}

\vspace{0.1cm}

We thank R.~Wilson for helping with proofreading.
G.~R.~was supported by the Funda\c{c}\~ao para
a Ci\^encia e a Tecnologia under Grant
No.~SFRH/BPD/26886/2006.
The authors also would like to thank the
International Institute of Physics,
Federal University of Rio Grande do
Norte, Brazil, where the revision of the manuscript was completed.
This work was supported
by Portuguese national funds through
FCT -- Funda\c{c}\~ao para a Ci\^encia e a Tecnologia,
under Grant No.~PTDC/FIS/113940/2009,
``Hadron Structure with Relativistic Models'',
and Project No.~PEst-OE/FIS/UI0777/2011.
This work was also supported partially by the European Union
under the HadronPhysics3 Grant No.~283286,
(K.~T.~), the University of Adelaide, and
the Australian Research Council through
Grant No.~FL0992247 (Anthony William Thomas).

%
%

\appendix

\section{Decuplet self energy}
\label{appMasses}

In this appendix we describe the formalism
on the decuplet self-energy and its
relation to the decuplet baryon masses.

The decuplet baryon $B'$ mass can be decomposed as
$M_{B'}= M_{0B'} + \Sigma_0^* (M_{B'})$, where
$\Sigma_0^*$ is the baryon self-energy at the pole position.
Considering only the pion cloud excitations,
we can represent the self-energy as
$\Sigma_0^* = G_{1B'} {\cal B}_1 + G_{2B'} {\cal B}_2$,
where ${\cal B}_1$ and ${\cal B}_2$ are
the value of the Feynman integrals, respectively,
with an intermediate octet and decuplet baryons,
where the factors $G_{1B}$ and $G_{2B}$ are the
coupling factors for the corresponding pion loops.
Using the couplings in Table~\ref{tableCouplings}
we obtain the factors listed
in Table~\ref{tableFlavor}.
As there are no pion cloud contributions
for the $\Omega^-$ baryon, $M_\Omega=M_{0B'}$.

\begin{table}
\begin{minipage}{3in}
\begin{tabular}{l c c}
\hline
\hline
$\pi B B'\qquad$ & ${\cal O}_{\pi B B'}$ & $\qquad g_{\pi B B'}\qquad$ \\
\hline
$\pi N N $ &  $g_{\pi NN} (\xi^*_\pi \cdot \mbox{\boldmath$\tau$}) $    & $g$     \\
$\pi \Lambda \Sigma$ & $g_{\pi \Lambda \Sigma} (\xi^*_\pi \cdot \xi_\Sigma)$
&  $\sfrac{2\sqrt{3}}{5}g $ \\
$\pi \Sigma \Sigma$ & $g_{\pi \Sigma \Sigma} (\xi^* \cdot {\bf J})$ &
$\sfrac{4}{5}g$ \\
$\pi \Xi \Xi$ & $g_{\pi \Xi \Xi}(\xi^*_\pi \cdot \mbox{\boldmath$\tau$}) $&
$- \sfrac{1}{5} g$ \\ [0.05in]
\hline
$\pi N \Delta$ & $g_{_{\pi N \Delta}} (\xi^*_\pi\cdot {\bf T})$ &
$\sfrac{2 \sqrt{2}}{5} g$ \\[0.05in]
$\pi\Lambda\Sigma^*$ &  $g_{_{\pi \Lambda \Sigma^*}}
(\xi^*_\pi\cdot\xi_{\Sigma^*})$  &
$\sfrac{2}{5} g$
\\[0.05in]
$\pi\Sigma\Sigma^*$ &  $g_{_{\pi \Sigma\Sigma^*}} (\xi^*_\pi\cdot{\bf J})$ &
$\sfrac{2 \sqrt{6}}{15} g$ \\[0.05in]
$\pi \Xi \Xi^*$ &  $g_{_{\pi \Xi \Xi^*}} (\xi^*_\pi\cdot \mbox{\boldmath$\tau$})$&
$\sfrac{2}{5} g$
\\[0.05in]
\hline
$\pi \Delta \Delta$ & $g_{_{\pi \Delta \Delta}} (\xi^*_\pi\cdot {\bf t})$ &
$g$ \\[0.05in]
$\pi\Sigma^* \Sigma^*$ &
$g_{_{\pi \Sigma^* \Sigma^*}} (\xi^*_\pi\cdot{\bf J})$ &
 $\sfrac{2 \sqrt{2}}{\sqrt{15}} g$ \\[0.05in]
$\pi \Xi^* \Xi^*$ &  $g_{_{\pi \Xi^* \Xi^*}} (\xi^*_\pi\cdot \mbox{\boldmath$\tau$})$&
$\sfrac{1}{\sqrt{5}} g$
\\[0.05in]
\hline
\hline
\end{tabular}
\caption{Pion-baryon couplings in $SU(3)$ symmetry with $\alpha \equiv D/(F+D) = 0.6$.
Here $\xi_\pi$ and $\xi_\Sigma$ are the isospin-1 polarization
vectors of the $\pi$ and $\Sigma$, \mbox{\boldmath$\tau$} are the isospin-1/2 matrices,
${\bf T}$ are the isospin 1/2 to 3/2 transition operator matrices,
${\bf J}$ are the isospin-1 matrices,
and ${\bf t}$ are the isospin 3/2 matrices.
For the diagonal operators, the isospin wave functions
of the initial and final baryons are suppressed. }
\label{tableCouplings}
\end{minipage}
\end{table}

\begin{table*}
\begin{tabular}{l l l}
\hline
\hline
$B\quad$ & $g^2 G_{1 B}$ &  $g^2G_{2B}$\\
\hline
$\Delta$ & $g_{\pi N\Delta}^2
\sum_\lambda(\xi_{\pi\lambda}\cdot T^\dagger)
(\xi^*_{\pi\lambda}\cdot T)=  \frac{8}{25} g^2 $  &
 $g_{\pi \Delta \Delta}^2\sum_\lambda(\xi_{\pi\lambda}\cdot \tau )
(\xi^*_{\pi\lambda}\cdot \tau)= g^2 $  \\
$\Sigma^* $ &  $g_{\pi \Sigma \Sigma^*}^2
\sum_\lambda(\xi_{\pi\lambda}\cdot{\bf J})(\xi^*_{\pi\lambda}\cdot{\bf J})+$ &
 $g_{\pi \Sigma^* \Sigma^*}^2
\sum_\lambda(\xi_{\pi\lambda}\cdot{\bf J})(\xi^*_{\pi\lambda}\cdot{\bf J})=
\frac{8}{15} g^2$
\\&
$g_{\pi \Lambda \Sigma^*}^2\sum_\lambda(\xi_{\pi\lambda}\cdot\xi_{\Sigma^*\mu})(\xi^*_{\pi\lambda}\cdot\xi^*_{\Sigma^* \mu})
=\frac{4}{15} g^2$   & \\[0.05in]
$\Xi^*$ &  $g_{\Xi \Xi^*}^2\sum_\lambda(\xi_{\pi\lambda}\cdot T^\dagger)(\xi^*_{\pi\lambda}\cdot T)= \sfrac{4}{25} g^2$ &
 $g_{\Xi^* \Xi^*}^2\sum_\lambda(\xi_{\pi\lambda}\cdot\tau)(\xi^*_{\pi\lambda}\cdot\tau)= \sfrac{1}{5} g^2$ \\
\hline
\hline
\end{tabular}
\caption{One pion-loop contributions to the
decuplet baryon self-energies with $\alpha=0.6$.}
\label{tableFlavor}
\end{table*}

\section{Overlap integral}
\label{appIntegral}

In this appendix we discuss the properties
of the integral ${\cal I}(Q^2)$ given by Eq.~(\ref{eqInt}),
also called {\it Body integral} \cite{NDelta},
for $Q^2=0$.
The value of ${\cal I}(0)$ measures the degree
of superposition of the radial wave functions
$\psi_{B'}$ and $\psi_B$, when $Q^2=0$.

In this appendix we show that
\ba
{\cal I}(0) \le 1,
\label{eqInt1}
\ea
where the equality holds only for the case $M_{B'}= M_B$.

Even in the equal mass case it is not assured
that ${\cal I}(0)=1$, unless $\psi_{B'} \equiv \psi_B$.
We can expect however, in the equal mass case,
${\cal I}(0) \simeq 1$, if the
two radial functions are very similar.

Next, we explicitly demonstrate
the relation~(\ref{eqInt1}).
In Sec.~\ref{appPart1} (Part 1) we explain the basic steps
of the demonstration, while in Sec.~\ref{appPart2} (Part 2),
we present the more technical details.

\subsection{Part 1}
\label{appPart1}

The integral ${\cal I}(0)$ is covariant,
therefore the result is independent
of the frame.
For simplicity, we use the $B'$ rest frame.

In the $B'$ rest frame, one can write
the initial ($P_-$) and final ($P_+$) momenta,
choosing $z$ as the photon direction
(momentum ${\bf q}$) as,
\ba
& &P_+= (M_{B'},0,0,0), \nonumber \\
& &P_-= (E_B,0,0,-|{\bf q}|), \nonumber \\
& &q = (\omega, 0,0, |{\bf q}|),
\ea
where $E_B=\sqrt{M_B^2+ |{\bf q}|^2}$ and $\omega$ are
the energies of the baryon $B$ and the photon,
respectively.

For $Q^2=0$, one has
\ba
\omega= |{\bf q}|= \frac{M_{B'}^2-M_B^2}{2 M_{B'}}.
\ea

Consider now the integral,
\ba
{\cal I}(0) = \int_k
\psi_{B'}(P_+,k) \psi_B(P_-,k).
\label{eqInt2}
\ea
As explained in Sec.~\ref{secValence}, the
radial wave functions are represented in terms
of the variables $\chi_{B'}$ and $\chi_{B}$,
defined by Eq.~(\ref{eqCHI}).
We can rewrite $\chi_{{B^*}}$ in terms
of a new variable $\eta_{B^*}$ defined by,
\ba
\eta_{B^*}\equiv \frac{P_{B^*} \cdot k}{M_{B^*} m_D},
\ea
where $B^*$ holds for $B$ or $B'$.
Then we can write,
\ba
\chi_{B^*}= 2 (\eta_{B^*} -1).
\ea

Redefining the diquark momentum ${\bf k}$ as
\mbox{\boldmath$\kappa$} $\equiv \sfrac{{\bf k}}{m_D}$, and
the diquark energy as $E_\kappa \equiv \sfrac{E_D}{m_D}$,
we can write,
\ba
& &
\eta_{B'} = E_{\kappa} = \eta_0, \\
& &
\eta_B = \tilde E_B   E_{\kappa} + q_B \kappa_z,
\label{eqEtaB}
\ea
where $\tilde E_B= \sfrac{E_B}{M_B}$,
$q_B= \frac{|{\bf q}|}{M_B}$ and $\eta_0 \equiv E_\kappa$.

With the above notations, we can write (\ref{eqInt2}) as,
\ba
{\cal I}(0) = \intKP
\psi_{B'}(\eta_{B'}) \psi_B(\eta_B).
\label{eqInt2a}
\ea
The normalization conditions in the same notations are,
\ba
& &
\intKP [\psi_{B}(\eta_0)]^2= 1,
\label{eqNorma1}\\
& &
\intKP [\psi_{B'}(\eta_0)]^2= 1.
\label{eqNorma2}
\ea
Note that, the both conditions are represented
in terms of the same argument $\eta_0$,
since in the rest frame of each particle,
all particles have the same value for $\eta_{B^\ast}$.

As shown in Sec.~\ref{appPart2},
we can prove that,
\ba
{\cal I}(0) &=& \intKP
\psi_{B'}(\eta_0) \psi_B(\eta_B), \nonumber \\
& \le &  \intKP
\psi_{B'}(\eta_0) \psi_B(\eta_0),
\label{eqInt3}
\ea
where the equality holds only for the case $M_{B'}= M_B$.
Then, using the Cauchy-Schwarz-H\"{o}lder inequality
for non-negative functions
\ba
\left[
 \intKP \psi_{B'}(\eta_0) \psi_B(\eta_0)\right]^2
\le
\left[
 \intKP [\psi_{B'}(\eta_0)]^2
\right]
\left[
 \intKP [\psi_{B}(\eta_0)]^2
\right],
\nonumber \\
\ea
we conclude from
Eqs.~(\ref{eqNorma1}) and~(\ref{eqNorma2}) that
\ba
\left|\intKP \psi_{B'}(\eta_0) \psi_B(\eta_0) \right| \le 1.
\label{eqCS2}
\ea

Combining the result~(\ref{eqCS2})
with (\ref{eqInt3})
for the case where both radial
wave functions are positive,
one has
\ba
{\cal I}(0) \le 1,
\ea
where the equality holds only for the case $M_{B'}=M_B$
[when $\eta_{B'} \equiv \eta_B = \eta_0$].

The details of the demonstration of Eq.~(\ref{eqInt3})
are in the next section.

\subsection{Part 2}
\label{appPart2}

Here we demonstrate the result given by Eq.~(\ref{eqInt3}).

Consider the integral,
\ba
{\cal I}(0) &=& \intKP
\psi_{B'}(\eta_0) \psi_B(\eta_B),
\label{eqInt3a}
\ea
where
\ba
\eta_B = \tilde E_B   E_{\kappa} + q_B \kappa_z.
\ea

Note that, in the case $M_{B'}=M_B$ one has
$q_B=0$, thus
$\eta_B = \eta_0$, and
\ba
{\cal I}(0)=
\intKP \psi_{B'}(\eta_0) \psi_B(\eta_0).
\ea

Consider now the case $q_B > 0$ (when $M_{B'} > M_B$).
In this case, according to Eq.~(\ref{eqEtaB}),
$\eta_B$ has an angular dependence.
Changing the integration variables to
$\kappa= \sfrac{|{\bf k}|}{m_D}$
and $z=\cos \theta$,
where $\theta$ is the angle with ${\bf q}$
($z$ direction), using $\kappa_z = \kappa z$,
one can represent Eq.~(\ref{eqInt3a}) as,
\ba
{\cal I}(0)=
m_D^2 \int_{0}^{+ \infty} \frac{\kappa^2 d \kappa}{(2\pi)^2 2 E_\kappa}
\psi_{B'}(\eta_0) \left[
\int_{-1}^1 dz \psi_{B}(\eta_B) \right].
\nonumber \\
\ea
Taking in consideration the
definition of $\psi_B$ given by Eqs.~(\ref{eqPsiN})-(\ref{eqPsiX}),
we can write $\psi_B$ as,
\ba
\psi_B(\eta_B)=
\frac{N_B}{4m_D} \,
\frac{1}{\alpha_i + \eta_B} \, \frac{1}{\alpha_j + \eta_B},
\label{eqPsiB2}
\ea
where $i,j=1,2,3,4$, but $i \ne j$, represent  the
possible indices,
and $\alpha_i = \sfrac{1}{2}(\beta_i -2)$ with $\alpha_i > -1$
(because $\beta_i > 0$).
Using the form (\ref{eqPsiB2}), we can now write,
\ba
{\cal I}(0) =
\frac{m_D N_B}{4}
\int_{0}^{+ \infty} \frac{\kappa^2 d \kappa}{(2\pi)^2 2 E_\kappa}
\psi_{B'}(\eta_0) I_z(q_B),
\ea
where
\ba
 I_z(q_B)=
\int_{-1}^1 dz
\frac{1}{\alpha_i + \eta_B}\frac{1}{\alpha_j + \eta_B}.
\label{eqIz1}
\ea
The last function includes all the $q_B$ dependence
of the integral ${\cal I}(0)$.
The $\kappa$ dependence on $I_z(q_B)$ is omitted for simplicity.

For the present case we can assume that $\alpha_j > \alpha_i$
without loss of generality.
Then, the integration in $z$ in~(\ref{eqIz1}) can
be performed with the decomposition:
\ba
I_z(q_B) =
\frac{1}{\alpha_j -\alpha_i}
\left[
\int_{-1}^1 dz
\frac{1}{\alpha_i + \eta_B} -
\int_{-1}^1 dz
\frac{1}{\alpha_j + \eta_B}
\right]. \nonumber \\
\ea
Defining
\ba
G(\alpha_i,q_B) &\equiv &
\int_{-1}^1 dz
\frac{1}{\alpha_i + \eta_B},  \nonumber \\
&=&
\frac{1}{q_B \kappa}
\log \frac{\alpha_i + \tilde E_B E_\kappa + q_B \kappa}{
\alpha_i + \tilde E_B E_\kappa - q_B \kappa},
\label{eqDefG}
\ea
we can write
\ba
I_z(q_B)=
\frac{1}{\alpha_j -\alpha_i}
\left[
G(\alpha_i,q_B) - G(\alpha_j,q_B)
\right].
\ea

The next step is to prove that $I_z(q_B)$
decreases when $q_B$ increases.
Performing the derivation in $q_B$,
one has,
\ba
\frac{d I_z}{d q_B}&=&
\frac{1}{\alpha_j -\alpha_i}
\Big\{
- \frac{1}{q_B}
\left[G(\alpha_i,q_B) - G(\alpha_j,q_B) \right] 
\nonumber \\
& &
+
\left[
H(\alpha_i,q_B) - H(\alpha_j,q_B)
\right] \frac{}{} \Big\},
\ea
where
\ba
\hspace{-.8cm}
& &
H(\alpha_i,q_B)=
\frac{1}{q_B \kappa}
\nonumber \\
\hspace{-.8cm}
& &
\times
\left[
\frac{1}{\alpha_i + \tilde E_B E_\kappa + q_B \kappa}-
\frac{1}{\alpha_i + \tilde E_B E_\kappa - q_B \kappa}
\right].
\label{eqDefH}
\ea

Let us consider first the term
\ba
T_1= G(\alpha_i,q_B) - G(\alpha_j,q_B).
\ea
Using the explicit form given by Eq.~(\ref{eqDefG}),
we can write,
\ba
T_1= \frac{1}{q_B \kappa}
\log
\frac{\alpha_i \alpha_j + t + (\alpha_i + \alpha_j)\tilde E_B E_\kappa + (\alpha_j -\alpha_i)q_B \kappa}{\alpha_i \alpha_j + t + (\alpha_i + \alpha_j)\tilde E_B E_\kappa - (\alpha_j -\alpha_i)q_B \kappa}, \nonumber
\\
\ea
where
$t = 1+ \kappa^2 + q_B^2$.
When $\alpha_j > \alpha_i$, the argument of the log function, $u$,
is larger than 1. Therefore $\log u > 0$
and
\ba
T_1 > 0,
\label{eqT1}
\ea
when  $\alpha_j > \alpha_i$.

Consider now
\ba
T_2= H(\alpha_i,q_L) - H(\alpha_j,q_L).
\ea
Working with the expression (\ref{eqDefH}), we obtain,
\ba
H(\alpha_i,q_L)= -\frac{1}{\left(\alpha_i + \tilde E_B E_\kappa\right)^2
- q_B^2 \kappa^2}.
\ea
Therefore,
\ba
T_2 =
\frac{1}{\left(\alpha_j + \tilde E_B E_\kappa\right)^2 - q_B^2 \kappa^2}-
\frac{1}{\left(\alpha_i + \tilde E_B E_\kappa\right)^2 - q_B^2 \kappa^2}.
\nonumber \\
\ea
If $\alpha_j > \alpha_i$,  one has
\ba
T_2 < 0.
\label{eqT2}
\ea

Combining the results (\ref{eqT1}) and (\ref{eqT2})
we conclude that,
\ba
\frac{d I_z}{d q_B} (q_B) <0,
\ea
when $\alpha_j > \alpha_i$.
As a consequence $I_z(q_B)$ decreases with increasing $q_B$
for $q_B > 0$, and
\ba
I_z(q_B) \le I_z(0),
\ea
where the equality holds only when $q_B=0$.

As $I_z(q_B)$ includes only the $q_B$ dependence
in ${\cal I}(0)$, we conclude also that
${\cal I}(0)$ is a decreasing function of $q_B$.
Furthermore, since ${\cal I}(0)$ is a continuous function
of $q_B$,
the maximum value for ${\cal I}(0)$ is obtained for the minimum
value of $q_B$, the case $q_B=0$,
when $\eta_B= \eta_0$.
Therefore,
\ba
{\cal I}(0) < \intKP \psi_{B'}(\eta_0) \psi_B(\eta_0),
\ea
if $q_B > 0$,
and
\ba
{\cal I}(0) = \intKP \psi_{B'}(\eta_0) \psi_B(\eta_0),
\ea
if $q_B=0$.

\section{Pion Cloud Dressing}
\label{appPionCloud}

In this appendix we present the expressions for the
pion cloud contributions.
We assume that the leading contribution
for the pion cloud dressing is given by the diagram
with a direct coupling of a photon to pion.
We assume also that in the first approximation,
the pion baryon vertex can be represented
by the results of the cloudy bag model
(CBM)~\cite{Thomas84,Thomas83,TsushimaCBM},
with the couplings determined by $SU(6)$ symmetry.
A similar approximation was also used in Ref.~\cite{Delta1600}.

In this description the pion cloud contributions
for the magnetic transition form factors
are determined by a function $F_{BB'}$, where
$B$ ($B'$) stands for the initial (final) state baryon.

Note that, the function $F_{BB'}$ can be a sum
of different amplitudes associated
with the several intermediate baryon $B_1$ states
(octet or decuplet baryons) as shown in Fig.~\ref{figPionCloud}.
Taking into account the possible spin and flavor
states, we can reduce the function $F_{BB'}(B_1)$
to a combination of scalar integral $H_{BB'}(B_1)$.

The results of the pion cloud contributions are presented
in Table~\ref{tableFBBp}.
We note that the analysis can be extended
to the kaon and $\eta$-meson clouds,
however, these meson contributions are known to be
smaller than those of the pion~\cite{TsushimaCBM}, and
thus we consider only the processes with the pion loops
in this study.

Finally, under an $SU(3)$ symmetry,
where all the octet members have a unique mass $M_B$
and all the decuplet members have a unique mass $M_{B'}$,
we can replace $H_{B B'}(B_1)$ by one single
function $H$ for all the cases of the $\gamma^* B \to B'$ reactions.
The results for this symmetry limit
are presented in Table~\ref{tableFBBp2}.
In this case, all the functions $F_{BB'}$
can be expressed in terms of the result
for the $\gamma^\ast N \to \Delta$
case ($F_{N\Delta}$),
as shown in the last column of  Table~\ref{tableFBBp2}.

\begin{table*}[t]
\begin{center}
\begin{tabular}{l     c }
\hline
\hline
    & $F_{BB'}$ \\
\hline
$\gamma^* N \to \Delta$  &
$F_{N\Delta} = \sfrac{4}{15\sqrt{3}}\left[
H_{N\Delta}(N) + 5 H_{N\Delta}(\Delta)\right] $  \\ [0.05in]
$\gamma^\ast \Lambda \to \Sigma^*$  &
$F_{\Lambda \Sigma^*} = \sfrac{8}{75}\left[
H_{\Lambda \Sigma^*}(\Sigma) + 5 H_{\Lambda\Sigma^*}(\Sigma^*)\right] $
 \\ [0.05in]
$\gamma^* \Sigma \to \Sigma^*$
& $\qquad F_{\Sigma \Sigma^*} = \sfrac{4}{75\sqrt{3}}
\left[ 3 H_{\Sigma \Sigma^*} (\Lambda) - 2 H_{\Sigma \Sigma^*} (\Sigma) + 5
H_{\Sigma \Sigma^*}(\Sigma^*) \right] J_3$
\\
 [0.05in]
$\gamma^* \Xi \to \Xi^*$ &
$F_{\Xi \Xi^*}= \sfrac{4}{75 \sqrt{3}}
\left[ H_{\Xi \Xi^*}(\Xi) + 5   H_{\Xi \Xi^*}(\Xi^*)
\right] \tau_3$
 \\
\hline
\hline
\end{tabular}
\end{center}
\caption{
Pion cloud contributions for $G_M^*$,
expressed in terms of the function $F_{BB'}$
(combination of the integrals $H_{BB'}(B_1)$).
}
\label{tableFBBp}
\end{table*}

\begin{table}[t]
\begin{center}
\begin{tabular}{l     c c }
\hline
\hline
    & $\qquad F_{BB'}(H) \qquad$ & $F_{BB'}(F_{N\Delta})$\\
\hline
$\gamma^* N \to \Delta$  &
$\frac{8}{5\sqrt{3}} H$ & $F_{N \Delta} $  \\ [0.05in]
$\gamma^\ast \Lambda \to \Sigma^*$  &
$ \frac{16}{25} H $
& $\frac{2 \sqrt{3}}{5} F_{N\Delta}$
 \\ [0.05in]
$\gamma^* \Sigma \to \Sigma^*$ &
$\frac{8}{25\sqrt{3}} H J_3 $
& $ \frac{1}{5} F_{N \Delta} J_3$
\\
 [0.05in]
$\gamma^* \Xi \to \Xi^*$ & $\frac{8}{25\sqrt{3}} H \tau_3$ &
$ \frac{1}{5} F_{N\Delta} \tau_3$
 \\
\hline
\hline
\end{tabular}
\end{center}
\caption{
Function $F_{BB'}$
in the $SU(3)$ limit $H_{BB'}(B_1) = H$.
The last column gives the result in terms
of that for $F_{N \Delta}$.}
\label{tableFBBp2}
\end{table}


\begin{thebibliography}{00}




\bibitem{Pascalutsa07}
  V.~Pascalutsa, M.~Vanderhaeghen and S.~N.~Yang,
  Phys.\ Rept.\  {\bf 437}, 125 (2007)
  [hep-ph/0609004].


\bibitem{Burkert04a}
V.~D.~Burkert and T.~S.~H.~Lee,
  Int.\ J.\ Mod.\ Phys.\ E {\bf 13}, 1035 (2004).


\bibitem{NSTAR}
  I.~G.~Aznauryan, A.~Bashir, V.~Braun, S.~J.~Brodsky, V.~D.~Burkert,
  L.~Chang, C.~.Chen and B.~El-Bennich {\it et al.},
   Int.\  J.\  Mod.\  Phys.\  E,\  {\bf 22}, 1330015 (2013)
  [arXiv:1212.4891 [nucl-th]].





\bibitem{PDG}
  K.~Nakamura {\it et al.}  [Particle Data Group],
  J.\ Phys.\ G {\bf 37}, 075021 (2010).



\bibitem{Keller11a}
  D.~Keller {\it et al.}  [CLAS Collaboration],
  Phys.\ Rev.\ D {\bf 83}, 072004 (2011)
  [arXiv:1103.5701 [nucl-ex]].

\bibitem{Keller11b}
  D.~Keller {\it et al.}  [CLAS Collaboration],
  Phys.\ Rev.\ D {\bf 85}, 059903 (2012)
  [arXiv:1111.5444 [nucl-ex]].


\bibitem{Taylor05}
  S.~Taylor {\it et al.}  [CLAS Collaboration],
  Phys.\ Rev.\ C {\bf 71}, 054609 (2005)
  [Erratum-ibid.\ C {\bf 72}, 039902 (2005)]
  [hep-ex/0503014].


\bibitem{Molchanov04}
  V.~V.~Molchanov {\it et al.}  [SELEX Collaboration],
  Phys.\ Lett.\ B {\bf 590}, 161 (2004)
  [hep-ex/0402026].









\bibitem{Lipkin73}
  H.~J.~Lipkin,
  Phys.\ Rev.\ D {\bf 7}, 846 (1973).


\bibitem{Koniuk80}
  R.~Koniuk and N.~Isgur,
  Phys.\ Rev.\ D {\bf 21}, 1868 (1980)
  [Erratum-ibid.\ D {\bf 23}, 818 (1981)].


\bibitem{Darewych83}
  J.~W.~Darewych, M.~Horbatsch and R.~Koniuk,
  Phys.\ Rev.\  D {\bf 28}, 1125 (1983).



\bibitem{Kaxiras85}
  E.~Kaxiras, E.~J.~Moniz and M.~Soyeur,
  Phys.\ Rev.\ D {\bf 32}, 695 (1985).



\bibitem{Warns}
  M.~Warns, H.~Schroder, W.~Pfeil and H.~Rollnik,
  Z.\ Phys.\ C {\bf 45}, 613 (1990);
  M.~Warns, W.~Pfeil and H.~Rollnik,
  Phys.\ Lett.\ B {\bf 258}, 431 (1991).




\bibitem{Sahoo95}
  R.~K.~Sahoo, A.~R.~Panda and A.~Nath,
  Phys.\ Rev.\  D {\bf 52}, 4099 (1995).


\bibitem{Wagner98}
  G.~Wagner, A.~J.~Buchmann and A.~Faessler,
  Phys.\ Rev.\ C {\bf 58}, 1745 (1998)
  [nucl-th/9808005].





\bibitem{Tiator04} 
  L.~Tiator, D.~Drechsel, S.~Kamalov, M.~M.~Giannini, E.~Santopinto and A.~Vassallo,
  Eur.\ Phys.\ J.\ A {\bf 19}, 55 (2004)
  [nucl-th/0310041].


\bibitem{Sharma10}
  N.~Sharma, H.~Dahiya, P.~K.~Chatley and M.~Gupta,
  Phys.\ Rev.\  D {\bf 81}, 073001 (2010)
  [arXiv:1003.4338 [hep-ph]].


\bibitem{Sharma13}
  N.~Sharma and H.~Dahiya,
  Pramana {\bf 80}, 237 (2013)
  [arXiv:1302.4167 [hep-ph]].



\bibitem{Santopinto12} 
  E.~Santopinto and M.~M.~Giannini,
  Phys.\ Rev.\ C {\bf 86}, 065202 (2012).


\bibitem{Schat96}
  C.~L.~Schat, C.~Gobbi and N.~N.~Scoccola,
  Phys.\ Lett.\ B {\bf 356}, 1 (1995)
  [hep-ph/9506227].


\bibitem{Abada96}
  A.~Abada, H.~Weigel and H.~Reinhardt,
  Phys.\ Lett.\ B {\bf 366}, 26 (1996)
  [hep-ph/9509396].

\bibitem{Haberichter97}
  T.~Haberichter, H.~Reinhardt, N.~N.~Scoccola and H.~Weigel,
  Nucl.\ Phys.\ A {\bf 615}, 291 (1997)
  [hep-ph/9610484].





\bibitem{Wang09}
  L.~Wang and F.~X.~Lee,
  Phys.\ Rev.\  D {\bf 80}, 034003 (2009)
  [arXiv:0905.1944 [hep-ph]].



\bibitem{Butler93a}
  M.~N.~Butler, M.~J.~Savage and R.~P.~Springer,
  Nucl.\ Phys.\ B {\bf 399}, 69 (1993)
  [hep-ph/9211247].

\bibitem{Butler93b}
  M.~N.~Butler, M.~J.~Savage and R.~P.~Springer,
  Phys.\ Lett.\  B {\bf 304}, 353 (1993)
  [arXiv:hep-ph/9302214].


\bibitem{Arndt04}
  D.~Arndt and B.~C.~Tiburzi,
  Phys.\ Rev.\ D {\bf 69}, 014501 (2004)
  [hep-lat/0309013].





\bibitem{Lebed11}
  R.~F.~Lebed and R.~H.~TerBeek,
  Phys.\ Rev.\ D {\bf 83}, 016009 (2011)
  [arXiv:1011.3237 [hep-ph]].





\bibitem{Bijker00}
  R.~Bijker, F.~Iachello and A.~Leviatan,
  Annals Phys.\  {\bf 284}, 89 (2000)
  [nucl-th/0004034].








\bibitem{Leinweber93}    
  D.~B.~Leinweber, T.~Draper and R.~M.~Woloshyn,
  Phys.\ Rev.\  D {\bf 48}, 2230 (1993)
  [arXiv:hep-lat/9212016].





\bibitem{Alexandrou05}
  C.~Alexandrou, Ph.~de Forcrand, H.~Neff, J.~W.~Negele,
  W.~Schroers and A.~Tsapalis,
  Phys.\ Rev.\ Lett.\  {\bf 94}, 021601 (2005)
  [arXiv:hep-lat/0409122];
  C.~Alexandrou, P.~.de Forcrand, T.~.Lippert, H.~Neff, J.~W.~Negele, 
  K.~Schilling, W.~Schroers and A.~Tsapalis,
  Phys.\ Rev.\ D {\bf 69}, 114506 (2004)
  [hep-lat/0307018].





\bibitem{Alexandrou08}
  C.~Alexandrou, G.~Koutsou, H.~Neff, J.~W.~Negele,
W.~Schroers and A.~Tsapalis,
  Phys.\ Rev.\  D {\bf 77}, 085012 (2008)
  [arXiv:0710.4621 [hep-lat]].










\bibitem{NDelta}
  G.~Ramalho, M.~T.~Pe\~na and F.~Gross,
  Eur.\ Phys.\ J.\  A {\bf 36}, 329 (2008)
  [arXiv:0803.3034 [hep-ph]].


\bibitem{NDeltaD}
  G.~Ramalho, M.~T.~Pe\~na and F.~Gross,
  Phys.\ Rev.\  D {\bf 78}, 114017 (2008)
  [arXiv:0810.4126 [hep-ph]].


\bibitem{OctetMedium}
  G.~Ramalho, K.~Tsushima and A.~W.~Thomas,
  J.\ Phys.\ G {\bf 40}, 015102 (2013)
  [arXiv:1206.2207 [hep-ph]].



\bibitem{OctetFF}
  G.~Ramalho and K.~Tsushima,
  Phys.\ Rev.\ D {\bf 84}, 054014 (2011)
  [arXiv:1107.1791 [hep-ph]].


\bibitem{Octet}
  F.~Gross, G.~Ramalho and K.~Tsushima,
  Phys.\ Lett.\  B {\bf 690}, 183 (2010)
  [arXiv:0910.2171 [hep-ph]].



\bibitem{Omega}
  G.~Ramalho, K.~Tsushima and F.~Gross,
  Phys.\ Rev.\  D {\bf 80}, 033004 (2009)
  [arXiv:0907.1060 [hep-ph]].




\bibitem{ExclusiveR}
  G.~Ramalho, F.~Gross, M.~T.~Pe\~na and K.~Tsushima, in
 {\it Proceedings of the 4th Workshop on Exclusive Reactions
at High Momentum Transfer},
  edited by A. Radyushkin
(World Scientific, Singapore, 2011), p.~287
  [arXiv:1008.0371 [hep-ph]].


\bibitem{Nucleon}
  F.~Gross, G.~Ramalho and M.~T.~Pe\~na,
  Phys.\ Rev.\  C {\bf 77}, 015202 (2008)
  [arXiv:nucl-th/0606029].

\bibitem{Nucleon2}
  F.~Gross, G.~Ramalho and M.~T.~Pe\~na,
  Phys.\ Rev.\ D {\bf 85}, 093005 (2012);
  F.~Gross, G.~Ramalho and M.~T.~Pe\~na,
  Phys.\ Rev.\ D {\bf 85}, 093006 (2012).



\bibitem{FixedAxis}
  F.~Gross, G.~Ramalho and M.~T.~Pe\~na,
  Phys.\ Rev.\  C {\bf 77}, 035203 (2008).




\bibitem{DeltaDFF}
  G.~Ramalho, M.~T.~Pe\~na and F.~Gross,
  Phys.\ Lett.\  B {\bf 678}, 355 (2009)
  [arXiv:0902.4212 [hep-ph]].

\bibitem{DeltaDFF2}
  G.~Ramalho, M.~T.~Pe\~na and F.~Gross,
  Phys.\ Rev.\  D {\bf 81}, 113011 (2010)
  [arXiv:1002.4170 [hep-ph]].


\bibitem{Deformation}
  G.~Ramalho, M.~T.~Pe\~na and A.~Stadler,
  Phys.\ Rev.\  D {\bf 86}, 86, 093022 (2012)
  [arXiv:1207.4392 [nucl-th]].


\bibitem{GE2Omega}
  G.~Ramalho and M.~T.~Pe\~na,
  Phys.\ Rev.\ D {\bf 83}, 054011 (2011)
  [arXiv:1012.2168 [hep-ph]].






\bibitem{LatticeD}
  G.~Ramalho and M.~T.~Pe\~na,
  Phys.\ Rev.\ D {\bf 80}, 013008 (2009)
  [arXiv:0901.4310 [hep-ph]].



\bibitem{Lattice}
  G.~Ramalho and M.~T.~Pe\~na,
  J.\ Phys.\ G {\bf 36}, 115011 (2009)
  [arXiv:0812.0187 [hep-ph]].


\bibitem{RoperS11}
  G.~Ramalho and K.~Tsushima,
  Phys.\ Rev.\ D {\bf 81}, 074020 (2010)
  [arXiv:1002.3386 [hep-ph]];
  G.~Ramalho and M.~T.~Pe\~na,
  Phys.\ Rev.\ D {\bf 84}, 033007 (2011)
  [arXiv:1105.2223 [hep-ph]].


\bibitem{Delta1600}
  G.~Ramalho and K.~Tsushima,
  Phys.\ Rev.\  D {\bf 82}, 073007 (2010)
  [arXiv:1008.3822 [hep-ph]];

  \bibitem{Various}
  G.~Ramalho and K.~Tsushima,
  Phys.\ Rev.\ D {\bf 86}, 114030 (2012)
  [arXiv:1210.7465 [hep-ph]]; 
  G.~Ramalho, D.~Jido and K.~Tsushima,
  Phys.\ Rev.\ D {\bf 85}, 093014 (2012)
  [arXiv:1202.2299 [hep-ph]].








\bibitem{Gross}
  F.~Gross,
  Phys.\ Rev.\  {\bf 186}, 1448 (1969);
  A.~Stadler, F.~Gross and M.~Frank,
  Phys.\ Rev.\  C {\bf 56}, 2396 (1997)
  [arXiv:nucl-th/9703043].









\bibitem{Boinepalli09}
  S.~Boinepalli, D.~B.~Leinweber, P.~J.~Moran, A.~G.~Williams, J.~M.~Zanotti and J.~B.~Zhang,
  Phys.\ Rev.\ D {\bf 80}, 054505 (2009)
  [arXiv:0902.4046 [hep-lat]].




\bibitem{Sanctis07} 
  M.~D.~Sanctis, M.~M.~Giannini, E.~Santopinto and A.~Vassallo,
  Phys.\ Rev.\ C {\bf 76}, 062201 (2007).




\bibitem{PDG2}
  C.~Amsler {\it et al.}  [Particle Data Group],
  Phys.\ Lett.\  B {\bf 667}, 1 (2008).
  [See page 1023 for information about the Naive Quark Model].



\bibitem{Chang11} 
  L.~Chang, Y.~-X.~Liu and C.~D.~Roberts,
  Phys.\ Rev.\ Lett.\  {\bf 106}, 072001 (2011)
  [arXiv:1009.3458 [nucl-th]].



\bibitem{Wilson12} 
  D.~J.~Wilson, I.~C.~Cloet, L.~Chang and C.~D.~Roberts,
  Phys.\ Rev.\ C {\bf 85}, 025205 (2012)
  [arXiv:1112.2212 [nucl-th]].



\bibitem{Cardarelli95} 
  F.~Cardarelli, E.~Pace, G.~Salme and S.~Simula,
  Phys.\ Lett.\ B {\bf 357}, 267 (1995)
  [nucl-th/9507037].


\bibitem{Ito95} 
  H.~Ito,
  Phys.\ Rev.\ C {\bf 52}, 1750 (1995).






\bibitem{Kelly98}
  J.~J.~Kelly,
  Phys.\ Rev.\  C {\bf 56}, 2672 (1997).


\bibitem{Batiz98}
  Z.~Batiz and F.~Gross,
  Phys.\ Rev.\  C {\bf 58}, 2963 (1998)
  [arXiv:nucl-th/9803053].





\bibitem{Jones73}
  H.~F.~Jones and M.~D.~Scadron,
  Annals Phys.\  {\bf 81}, 1 (1973).






\bibitem{Bartel68}
   W.~Bartel, B.~Dudelzak, H.~Krehbiel, J.~McElroy, U.~Meyer-Berkhout, W.~Schmidt, V.~Walther and G.~Weber,
  Phys.\ Lett.\ B {\bf 28}, 148 (1968).

\bibitem{Stein75}
   S.~Stein, W.~B.~Atwood, E.~D.~Bloom, R.~L.~Cottrell, H.~C.~DeStaebler, C.~L.~Jordan, H.~Piel and C.~Y.~Prescott {\it et al.},
  Phys.\ Rev.\ D {\bf 12}, 1884 (1975).







\bibitem{CLAS}
  K.~Joo {\it et al.}  [CLAS Collaboration],
  Phys.\ Rev.\ Lett.\  {\bf 88}, 122001 (2002)
  [arXiv:hep-ex/0110007];
  I.~G.~Aznauryan {\it et al.}  [CLAS Collaboration],
  Phys.\ Rev.\ C {\bf 80}, 055203 (2009)
  [arXiv:0909.2349 [nucl-ex]].






\bibitem{Tiator01}
  L.~Tiator, D.~Drechsel, O.~Hanstein, S.~S.~Kamalov and S.~N.~Yang,
  Nucl.\ Phys.\  A {\bf 689}, 205 (2001)
  [arXiv:nucl-th/0012046].




\bibitem{Drechsel07}
  D.~Drechsel, S.~S.~Kamalov and L.~Tiator,
  Eur.\ Phys.\ J.\ A {\bf 34}, 69 (2007)
  [arXiv:0710.0306 [nucl-th]].






\bibitem{SatoLee}
  T.~Sato and T.~S.~H.~Lee,
  Phys.\ Rev.\  C {\bf 63}, 055201 (2001)
  [arXiv:nucl-th/0010025].




\bibitem{Diaz07a}
 B.~Julia-Diaz, T.~S.~H.~Lee, T.~Sato and L.~C.~Smith,
  Phys.\ Rev.\  C {\bf 75}, 015205 (2007).



\bibitem{Kamalov}
  S.~S.~Kamalov, S.~N.~Yang, D.~Drechsel, O.~Hanstein and L.~Tiator,
  Phys.\ Rev.\  C {\bf 64}, 032201 (2001)
  [arXiv:nucl-th/0006068];





\bibitem{Carlson}
  C.~E.~Carlson,
  Phys.\ Rev.\ D {\bf 34}, 2704 (1986);
  C.~E.~Carlson,
  eConf C {\bf 010430}, W09 (2001)
  [hep-ph/0106290].





\bibitem{Wolf90}
  G.~Wolf, G.~Batko, W.~Cassing, U.~Mosel, K.~Niita and M.~Schaefer,
  Nucl.\ Phys.\  A {\bf 517}, 615 (1990).




\bibitem{DeltaTL}
  G.~Ramalho and M.~T.~Pe\~na,
  Phys.\ Rev.\ D {\bf 85}, 113014 (2012)
  [arXiv:1205.2575 [hep-ph]].


\bibitem{Capstick95}
  S.~Capstick and B.~D.~Keister,
  Phys.\ Rev.\ D {\bf 51}, 3598 (1995)
  [nucl-th/9411016].







\bibitem{Thomas84}
  A.~W.~Thomas,
  Adv.\ Nucl.\ Phys.\  {\bf 13}, 1 (1984).

\bibitem{Thomas83}
  S.~Theberge and A.~W.~Thomas,
  Nucl.\ Phys.\  A {\bf 393}, 252 (1983).


\bibitem{TsushimaCBM}
  K.~Tsushima, T.~Yamaguchi, M.~Takizawa, Y.~Kohyama and K.~Kubodera,
  Phys.\ Lett.\  B {\bf 205}, 128 (1988);
  K.~Tsushima, T.~Yamaguchi, Y.~Kohyama and K.~Kubodera,
  Nucl.\ Phys.\  A {\bf 489}, 557 (1988);
  T.~Yamaguchi, K.~Tsushima, Y.~Kohyama and K.~Kubodera,
  Nucl.\ Phys.\  A {\bf 500}, 429 (1989).







\bibitem{CLAS2}
  V.~V.~Frolov {\it et al.},
  Phys.\ Rev.\ Lett.\  {\bf 82}, 45 (1999)
  [arXiv:hep-ex/9808024];
  M.~Ungaro {\it et al.}  [CLAS Collaboration],
  Phys.\ Rev.\ Lett.\  {\bf 97}, 112003 (2006)
  [arXiv:hep-ex/0606042].
  A.~N.~Villano, P.~Stoler, P.~E.~Bosted, S.~H.~Connell, M.~M.~Dalton, M.~K.~Jones, V.~Kubarovsky and G.~S.~Adams {\it et al.},
  Phys.\ Rev.\ C {\bf 80}, 035203 (2009)
  [arXiv:0906.2839 [nucl-ex]].








\end{thebibliography}
\end{document}